\documentclass[iop]{emulateapj}
\pdfoutput=1
\usepackage{amsmath,amssymb}
\usepackage{xspace}
\usepackage{graphicx}
\usepackage{epsfig}
\usepackage{adjustbox}
\def\one{\uppercase\expandafter{\romannumeral1}} 
\def\two{\uppercase\expandafter{\romannumeral2}} 
\def\thr{\uppercase\expandafter{\romannumeral3}} 
\def\four{\uppercase\expandafter{\romannumeral4}} 
 
\def\cdfs{\hbox{CDF-S}\ } 
\def\nxv{$\sigma_{\rm nxv}^2$\ }  
\def\nxvt{$\sigma_{\rm nxv}^2$}  
 
\def\mytorus{\textit{\rm {\footnotesize MYTorus}}}
\def\lsim{\mathrel{\rlap{\lower4pt\hbox{\hskip1pt$\sim$}}
    \raise1pt\hbox{$<$}}}                
\def\gsim{\mathrel{\rlap{\lower4pt\hbox{\hskip1pt$\sim$}}
    \raise1pt\hbox{$>$}}}                
\usepackage{hyperref} 
\usepackage{xcolor}

\slugcomment{}
\shorttitle{AGN X-ray variability in CDF-S}
\shortauthors{Zheng et~al.}

\begin{document}
\title{Deepest view of AGN X-ray variability with the 7 Ms Chandra Deep Field-South Survey}
\author{X.~C. Zheng\altaffilmark{1,2},
Y.~Q. Xue\altaffilmark{1,2},
W.~N. Brandt\altaffilmark{3,4,5},
J.~Y. Li\altaffilmark{1,2},
M.~Paolillo\altaffilmark{6,7,8},
G. Yang\altaffilmark{3,4},
S.~F. Zhu\altaffilmark{3,4},
B. Luo\altaffilmark{9,10,11},
M.~Y. Sun\altaffilmark{1,2},
T.~M. Hughes\altaffilmark{12,1,2,13},
F.~E. Bauer\altaffilmark{14,15,16},
F. Vito\altaffilmark{3,4},
J.~X. Wang\altaffilmark{1,2},
T. Liu\altaffilmark{1,2,17},
C. Vignali\altaffilmark{18,19}, and
X.~W. Shu\altaffilmark{20}
}
\altaffiltext{1}{CAS Key Laboratory for Research in Galaxies and Cosmology, Department of Astronomy, University of Science and Technology of China, Hefei 230026, China; isaac10@mail.ustc.edu.cn, xuey@ustc.edu.cn}
\altaffiltext{2}{School of Astronomy and Space Science, University of Science and Technology of China, Hefei 230026, China}
\altaffiltext{3}{Department of Astronomy and Astrophysics, 525 Davey Lab, The Pennsylvania State University, University Park, PA 16802, USA}
\altaffiltext{4}{Institute for Gravitation and the Cosmos, The Pennsylvania State University, University Park, PA 16802, USA}
\altaffiltext{5}{Department of Physics, 104 Davey Lab, The Pennsylvania State University, University Park, PA 16082, USA}
\altaffiltext{6}{Dipartimento di Fisica ``Ettore Pancini'', Universit\`{a} di Napoli Federico II, via Cintia, 80126, Italy}
\altaffiltext{7}{INFN - Unit\`{a} di Napoli, via Cintia 9, 80126, Napoli, Italy}
\altaffiltext{8}{Agenzia Spaziale Italiana-Science Data Center, Via del Politecnico snc, 00133, Roma, Italy}
\altaffiltext{9}{School of Astronomy and Space Science, Nanjing University, Nanjing 210093, China}
\altaffiltext{10}{Key Laboratory of Modern Astronomy and Astrophysics  (Nanjing University), Ministry of Education, Nanjing 210093, China}
\altaffiltext{11}{Collaborative Innovation Center of Modern Astronomy and Space Exploration, Nanjing 210093, China}
\altaffiltext{12}{Instituto de F\'{i}sica y Astronom\'{i}a, Universidad de Valpara\'{i}so, Avda. Gran Breta\~{n}a 1111, Valpara\'{i}so, Chile}
\altaffiltext{13}{Chinese Academy of Sciences South America Center for Astronomy,
China-Chile Joint Center for Astronomy, Camino El Observatorio \#1515,
Las Condes, Santiago, Chile}
\altaffiltext{14}{Instituto de Astrof{\'{\i}}sica and Centro de Astroingenier{\'{\i}}a, Facultad de F{\'{i}}sica, Pontificia Universidad Cat{\'{o}}lica de Chile, Casilla 306, Santiago 22, Chile}
\altaffiltext{15}{Millennium Institute of Astrophysics (MAS), Nuncio Monse{\~{n}}or S{\'{o}}tero Sanz 100, Providencia, Santiago, Chile}
\altaffiltext{16}{Space Science Institute, 4750 Walnut Street, Suite 205, Boulder, Colorado 80301}
\altaffiltext{17}{Astronomy Department, University of Massachusetts, Amherst, MA 01003, USA}
\altaffiltext{18}{Dipartimento di Fisica e Astronomia, Alma Mater Studiorum, Universit\`{a} degli Studi di Bologna, Viale Berti Pichat 6/2, 40127 Bologna, Italy}
\altaffiltext{19}{INAF -- Osservatorio Astronomico di Bologna, Via Ranzani 1, 40127 Bologna, Italy}%
\altaffiltext{20}{Department of Physics, Anhui Normal University, Wuhu, Anhui, 241000, China}
\begin{abstract}

We systematically analyze X-ray variability of active galactic nuclei (AGNs) in the 7~Ms \textit{Chandra} Deep Field-South survey. On the longest timescale ($\approx~17$ years), we find only weak (if any) dependence of X-ray variability amplitudes on energy bands or obscuration. We use four different power spectral density (PSD) models to fit the anti-correlation between normalized excess variance (\nxvt) and luminosity, and obtain a best-fit power law index $\beta=1.16^{+0.05}_{-0.05}$ for the low-frequency part of AGN PSD. We also divide the whole light curves into 4 epochs in order to inspect the dependence of \nxv on these timescales, finding an overall increasing trend. The analysis of these shorter light curves also infers a $\beta$ of $\sim 1.3$ that is consistent with the above-derived $\beta$, which is larger than the frequently-assumed value of $\beta=1$. We then investigate the evolution of \nxvt. No definitive conclusion is reached due to limited source statistics but, if present, the observed trend goes in the direction of decreasing AGN variability at fixed luminosity toward large redshifts.
We also search for transient events and find 6 notable candidate events with our considered criteria. Two of them may be a new type of fast transient events, one of which is reported here for the first time. We therefore estimate  a rate of fast outbursts $\langle\dot{N}\rangle = 1.0^{+1.1}_{-0.7}\times 10^{-3}~\rm galaxy^{-1}~yr^{-1}$ and
a tidal disruption event~(TDE) rate $\langle\dot{N}_{\rm TDE}\rangle=8.6^{+8.5}_{-4.9}\times 10^{-5}~\rm galaxy^{-1}~yr^{-1}$ assuming the other four long outbursts to be TDEs.

\end{abstract}


\section{Introduction}
Active galactic nuclei (AGNs) are among the most luminous objects in the universe and have violent activities. It is often believed that their energy comes from the accretion of matter onto super massive black holes (SMBHs) at galactic centers. At present, there remain many unanswered questions about AGN structure and how matter falls into them. Variability existing in all wavelengths is becoming an increasingly essential aspect to answer these questions. In particular, X-ray variability is of great importance because X-rays are radiated from the most inner part of the system. Rapid variability in X-rays can provide a unique view to understand black hole accretion physics and is an efficient way to search for moderate- and low-luminosity AGNs \citep[see, e.g.,][ and references therein]{Young12,Xue17}. 
 
With the help of high-quality monitoring \citep[e.g.,][]{Uttley05, McHardy07,Gonzalez12}, people are able to explore X-ray variability on different timescales using the power spectral density. It is found that the X-ray variability characteristics of AGNs are quite similar to those of X-ray black hole binary (BHB) candidates \citep[e.g.,][]{Cui97a,Cui97b,McHardy06}. The high-frequency part of an AGN PSD is often fitted by a power law with an index about 2 \citep[e.g.,][]{Zhou10,Gonzalez12,Kelly13}. High-quality longer observations reveal that the PSDs of some AGNs flatten below a break frequency and the index becomes about 1 \citep[e.g.,][]{Uttley02,Uttley05,McHardy06,Breedt09,Gonzalez12}. In at least one AGN, Ark 564 \citep{McHardy07}, even a second break could be seen, although a multiple Lorentzian model, which is usually adopted in BHB PSD fitting, might be a better choice in that case.

In previous studies \citep[e.g.,][]{Nandra97,Papadakis04,McHardy06,Zhou10,Ponti12}, evidence has accumulated that X-ray variability is correlated with physical properties of AGNs. Luminous AGNs tend to have relatively weak variability \citep[e.g.,][]{Nandra97,Ponti12}. The break frequency of the PSD has become an important parameter because of its potential correlation with black hole mass and accretion rate \citep[e.g.,][]{McHardy06,Gonzalez12}. Studies have shown that the correlation is essentially the same for BHBs and AGNs \citep[e.g.,][]{McHardy06,Kording07}, revealing that the accretion process is similar in both small and large accreting systems. Therefore, using the variability characteristics could help us explore the physics in the central black holes.

For long-term variability studies, using simpler methods rather than PSD, such as $\chi^2$ and normalized excess variance \nxvt, to assess variability significance and quantify variability amplitude are also routine \citep[e.g.,][]{Almaini00,Nikolajuk04,Paolillo04,Gonzalez11,Lanzuisi14,Yang16}, given that PSD measurements usually require high-quality continuous monitoring, which are only feasible for exploring short-timescale variability, or rely on the continuous-time autoregressive moving average \citep[CARMA; e.g.,][]{Kelly09,Kelly13,Kelly14,Simm16} model simulations. 
Previous studies \citep[e.g.,][]{Oneill05,Zhou10,Ponti12,Kelly13,Pan15} have established that there is a tight correlation between \nxv and black hole mass. Therefore, \nxv can be used to measure or at least constrain black hole mass of AGNs. However, long-term variability studies usually involve irregular sparse sampling, uneven exposure times, and low signal-to-noise ratios (S/N). These factors could introduce large uncertainties in the calculation of single-epoch \nxv. \citet{Allevato13} discussed this issue and determined how these factors might introduce biases and cause scatters. Ensemble excess variance, which is the average of the measurements from several epochs or similar sources, is commonly utilized to reduce the influence of these factors \citep[e.g.,][]{Lanzuisi14,Vagnetti16}.

Due to the limitation of instrumental sensitivity and observational strategy, most of the studied objects are local and bright in previous works. However, in recent years, aided by instrumental development and accumulation of deep X-ray survey data, people have become able to study AGN X-ray variability in the deeper universe with longer timescales \citep[e.g.,][]{Paolillo04,Paolillo17,Lanzuisi14,Yang16}.
A case in point is the \textit{Chandra} Deep Fields (CDFs), which consist of the 2~Ms \textit{Chandra} Deep Field-North \citep[CDF-N;][]{Brandt01,Alexander03,Xue16}, the 7~Ms \textit{Chandra} Deep Field-South \citep[CDF-S;][]{Giacconi02,Luo08a,Luo17,Xue11}, and the 250~ks Extended \textit{Chandra} Deep Field-South \citep[E-CDF-S;][]{Lehmer05,Xue16}. 
Together, these surveys allow us to probe low- and moderate-luminosity AGNs at $z\lsim 6$ with $\lsim 7$~Ms exposure in a timespan of $\lsim 17$ years \citep[see][for more details about the CDFs]{Xue17}. 

In particular, the 7~Ms \cdfs \citep[][hereafter L17]{Luo17} is the deepest and most sensitive X-ray survey even taken, providing an unprecedented sample of $\approx 1000$ X-ray sources ($\approx 71$\% being classified as AGNs) in the distant universe. Previously, \citet{Paolillo04} used the 1~Ms \cdfs data \citep{Giacconi02} to analyze AGN X-ray variability and studied the anti-correlation between variable amplitude and AGN luminosity; they also suggested that the relation might evolve when taking redshifts into account. \citet{Young12} measured the X-ray variability using the 4~Ms \cdfs data \citep{Xue11} to identify distant low-luminosity AGNs that are typically missed by other AGN-selection criteria. \citet{Yang16} investigated the photon flux, X-ray luminosity, and absorption variability of the brightest AGNs in the 6~Ms CDF-S, exploring the nature of long-term AGN X-ray variability. Most recently, \citet{Paolillo17} made use of the 7~Ms \cdfs data to examine X-ray variability, thus tracing the accretion history of SMBHs. 

In addition to long-term AGN X-ray variability, the 7~Ms \cdfs data could also be utilized to search for X-ray transient events, especially tidal disruption events (TDEs hereafter). A TDE occurs when a stray star is sufficiently close to a SMBH and thereby ripped off by its strong tidal force. 
Despite of many efforts, there are still discrepancies in the estimates of TDE rate between observational studies \citep[e.g.,][]{Donley02,Luo08b,van14} and theoretical works \citep[e.g.,][]{Wang04,Stone16}. 
We could benefit from the high sensitivity and long monitoring time ($\approx 17$~years) of the 7~Ms \cdfs data and obtain a simple estimate of TDE rate. Apart from that, there could also be some interesting transient events recorded in the 7~Ms data. For instance, L17 and \citet{Bauer17} have already found a likely new type of outburst event, whose exact nature remains a mystery \citep{Bauer17}. A systematic search may uncover additional possible outbursts of great interest.

In this paper, we carry out a systematic and robust study of AGN X-ray variability in the 7~Ms CDF-S by taking several biases into account, which focuses mainly on \nxv properties, PSD constraints, and a rough estimate of TDE rate, and aims to obtain an ultradeep and unbiased view of AGN X-ray variability. 
This paper is organized as follows. In Section~\ref{sec:data} we briefly introduce the 7~Ms \cdfs data. In Section~\ref{sec:lcextract} we present light curve extraction and initial sample construction. In Section~\ref{sec:anal} we describe how we calculate \nxv reliably and thus build an unbiased sample for subsequent investigations. In Section~\ref{sec:res} we perform \nxvt-related correlation analyses and compare different PSD models. In Section~\ref{sec:trans} we search for likely transient events. Finally, we conclude this paper with a brief summary of our results in Section~\ref{sec:sum}. Throughout this paper, we adopt a cosmology with $H_0=67.8~\rm km~s^{-1}~Mpc^{-1}$, $\Omega_{\rm M}=0.308$, and $\Omega_{\Lambda}=0.692$ \citep{Planck16}.

\section{Data}\label{sec:data}

We utilize the 7~Ms \cdfs data (L17) to study long-term AGN X-ray variability. The 7~Ms \cdfs consists of 102 observations performed by the Advanced CCD Imaging Spectrometer image array (ACIS-I) onboard \textit{Chandra} from October 1999 to March 2016 (thus covering a total timespan of $\sim 5.2\times10^8~\rm s$) with a total exposure time of nearly 7~Ms. CIAO v4.8 with CALDB v4.7.0 was adopted to process the data (see L17 for more details). A merged event list and exposure maps of individual observations in different energy bands were produced and used to extract light curves.

Our sample selection is based on the 7~Ms \cdfs main catalog (L17) that contains 1008 sources. L17 first produced a list of candidate sources that were detected by \texttt{WAVDETECT} \citep{Freeman02} with a false-positive probability threshold of $10^{-5}$, and then used \texttt{ACIS EXTRACT} \citep[AE;][]{Broos10} to extract photometry and compute binomial no-source probabilities ($P_{\rm B}$) to exclude low-significance candidates, thereby obtaining a more conservative (i.e., ${\rm P_B}<0.007$) source list as the main catalog. 

In order to investigate the connection between variability and spectral properties for the bright AGNs, we perform spectral fitting for sources with reliable \nxv measurements (see Section~\ref{sec:finsamp}) in the 7~Ms exposure using XSPEC \citep[version 12.9.0;][]{Arnaud96}. For each such source, we fit the unbinned source and background spectra simultaneously and adopt the Cash statistic to find the best-fit parameters. The background spectrum is fitted with the $cplinear$ model. The source spectrum is fitted by a combination of the background component and the commonly used source model $phabs \times (zwabs \times zpow + zgauss + constant \times zpow)$, which includes the intrinsic power law, Fe~K$\alpha$ emission line, and soft-excess component to obtain the intrinsic photon index $\Gamma$, intrinsic X-ray luminosity $L_{\rm X}$, and hydrogen column density $N_{\rm H}$. For highly obscured sources ($N_{\rm H}\gtrsim 10^{23}~\rm cm^{-2}$), we use the \mytorus~model \citep{Murphy09} instead to obtain more accurate parameter estimates. Details of this spectral fitting method will be presented in Li et al. (in prep). 

\section{Light curve extraction and initial sample construction}\label{sec:lcextract}
\subsection{Light curves}\label{sec:lc}

Most sources in the \cdfs have a very low count rate and S/N. To enhance S/N while retaining as many features in a light curve as possible, we decide to adopt a binning strategy such that each data point of the resulting light curve represents the binned result of an individual observation whose exposure time ranges from $\approx$30~ks to $\approx$150~ks. Although many sources are still too faint for reliable analysis given this binning scheme, the bright ones we focus on would have enough S/N for variability measurement. 

\begin{figure*}[tp]
\includegraphics[width=\textwidth]{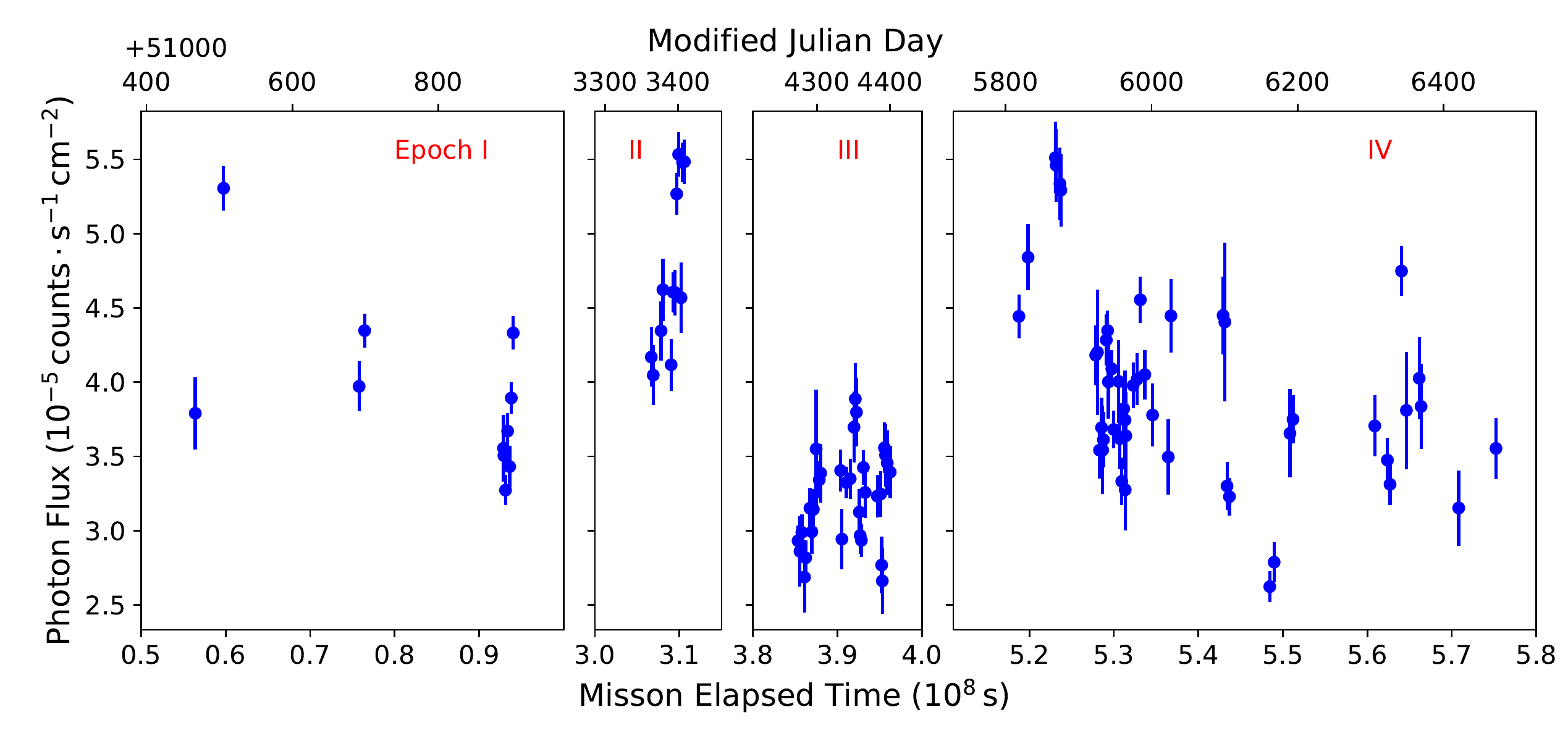}
\caption{0.5--7 keV light curve of the source with XID=495 in L17, which is the brightest source (having 56916.2 full-band net counts) in our sample; the time is shown as the Mission Elapsed Time of \textit{Chandra} (bottom $x$-axis) and the Modified Julian Day (top $x$-axis), respectively. Each data point of the light curve represents the binned result of an individual observation. The light curve is divided into four epochs.}
\label{fig:lcsamp}
\end{figure*}

In the light curve extraction procedure, there are complexities from instruments that would influence our results including vignetting, CCD gaps, bad pixels, and quantum efficiency degradation. Therefore, we adopt a similar solution to that of \citet{Young12}, using effective exposure maps to calibrate these instrumental effects. For each source, we calculate the 90\% encircled-energy fraction radii $R_{90}$ in every observation based on point-spread function modeling results in \citet{Xue11}. Then we use a circular region with a radius $R_{\rm src}$ to estimate source counts and an annulus region with an inner radius $R_{\rm bkg,in}$ and an outer radius $R_{\rm bkg,out}$ to estimate background counts ($R_{\rm src}$, $R_{\rm bkg,in}$ and $R_{\rm bkg,out}$ are listed in Table  \ref{tab:aper}). These aperture choices are made after trying a series of aperture combinations to maximize S/N of light curves. We only select events with grades 0, 2, 3, 4, and 6, and exclude those that also fall into the source area of another object. Finally, after background subtraction, we obtain our long-term light curves in three energy bands: 0.5--2 (soft band), 2--7 (hard band), and 0.5--7 keV (full band; in the observed frame). We present the full-band light curve of the brightest source as an example in Figure~\ref{fig:lcsamp}.

It should be noted that because we use the above simplified procedure instead of AE to extract the light curves, the amount of photon counts would be slightly different from that given by L17. For consistency, we adopt the total counts from our light-curve extractions in the following analysis.

\begin{table}[tp]
    \centering
        \caption{Aperture radii adopted in light curve extraction}
            \resizebox{\linewidth}{!}{
            \begin{tabular}{ccccc}\hline\hline    
            Net counts & Axis angle ($\arcmin$) & $R_{\rm src}/R_{90}$ & $R_{\rm bkg,in}/R_{90}$ & $R_{\rm bkg,out}/R_{90}$ \\\hline
            All& $<2$ & 1 & 1.2 & 7.5\\
            0--1000& $>2$ & 1 & 1.5 & 5\\
            1000--15000& $>2$ & 1.3 & 2 & 5\\
            $>15000$ & $>2$ & 1.7 & 2.5 & 5 \\ \hline
            \end{tabular}\label{tab:aper}
            }
\end{table}

We compute the errors of source and background counts using both the Gehrels approximation \citep{Gehrels86} and square root of counts as the following:
\begin{align}
    \Delta n_{\rm Geh} &= \frac{1}{2} (\Delta n_{\rm Geh,upper}+\Delta n_{\rm Geh,lower})\label{eq:geh} \\
    \Delta n_{\rm Geh,upper} &= 1+\sqrt{n+0.75} \\
    \Delta n_{\rm Geh,lower} &= n-n (1-\frac{1}{9n}-\frac{1}{3\sqrt{n}})^3 \\
    \Delta n_{\rm sqrt} &= \sqrt{n} \label{eq:sqerr}
\end{align}
The Gerhels approximation is a better error estimation in the low-counts regime, but the square root of counts are the standard deviation of Poisson distribution theoretically. These two approximations both have their respective advantages in following analyses (see more details in Section~\ref{sec:anal}).

As shown in Fig.~\ref{fig:lcsamp}, the 102 individual observations are roughly distributed in four periods with 1~Ms, 1~Ms, 2~Ms, and 3~Ms exposures, respectively. Therefore, we divide the long light curve into four parts that correspond to the four epochs. These four short light curves provide variability information of four different timescales of a source.

We also use another binning strategy in order to search for transient events in the \cdfs. Given that a TDE usually has a decay time of a few months to years, we rebin the data in bins of about 3 months to make a new light curve of a source (more details are provided in Section~\ref{sec:trans}).

\subsection{Initial sample construction}\label{sec:dred}

As mentioned above, many faint sources do not have enough counts for variability estimation. Furthermore, some sources were not covered by all the 102 observations. Inconsistent observing patterns could introduce large uncertainties in the following analysis. Therefore, we construct our initial sample based on the following criteria:
\begin{itemize}
 \item[1] The source was classified as an AGN in L17, but not classified as a radio-loud AGN in \citet{Bonzini13}.
 \item[2] The source has more than 100 full-band net counts in the 7~Ms exposure.
 \item[3] The overall length of the long light curve is larger than 15.2~years (i.e., $4.8\times 10^{8}\rm s$, $\sim 90\%$ of the longest light curve).
 \item[4] The source was covered by more than 70 observations.
 \item[5] The source region is outside $R_{\rm bkg,in}$ of any other sources.
\end{itemize}

As a result, 283 of the 1008 sources meet these initial requirements. However, it should be noted that the 100 counts cut is still not enough to discard all sources that are not suitable for reliable variability analyses. We intend to include as many sources as possible while ensuring that the variability estimation of these sources does not suffer from the uncertainties arising form low count rates. Therefore, we have to figure out what would happen when our measuring methods are used in the low-counts regime, in order to secure an unbiased sample (see Section~\ref{sec:anal} for details).

\section{Data analysis}\label{sec:anal}
\subsection{Normalized excess variance}\label{sec:nxv}

To quantify the variability amplitude of a light curve, we compute the normalized excess variance and its error \citep{Vaughan03} as the following:
\begin{equation}
\resizebox{\linewidth}{!}{$
    \sigma_{\rm nxv}^2 = \frac{1}{ (N_{\rm obs}-1)\langle\dot{n}\rangle^2}\sum\limits_{i=1}^{N_{\rm obs}}  (\dot{n}_i-\langle\dot{n}\rangle)^2- \frac{1}{N_{\rm obs}\langle\dot{n}\rangle^2}\sum\limits_{i=1}^{N_{\rm obs}}\sigma_{i,\rm err,var}^2$}
\end{equation}
\begin{equation}
err (\sigma_{\rm nxv}^2)=\sqrt{\frac{2}{N_{\rm obs}} (\frac{\overline{\sigma_{\rm err,var}^2}}{\langle\dot{n}\rangle^2})^2+\frac{\overline{\sigma_{\rm err,var}^2}}{N_{\rm obs}}\frac{4\sigma_{\rm nxv}^2}{\langle\dot{n}\rangle^2}} \label{eq:verr}
\end{equation}
where $N_{\rm obs}$ is the number of observations, $\dot{n}_i$ and $\sigma_{i,\rm err,var}$ are the photon flux and its error of the source in the $i$th observation, and $\langle \dot{n}\rangle$ is the exposure-weighted average photon flux of the light curve.

\begin{figure}[tp]
\centering
\includegraphics[width=1.1\linewidth]{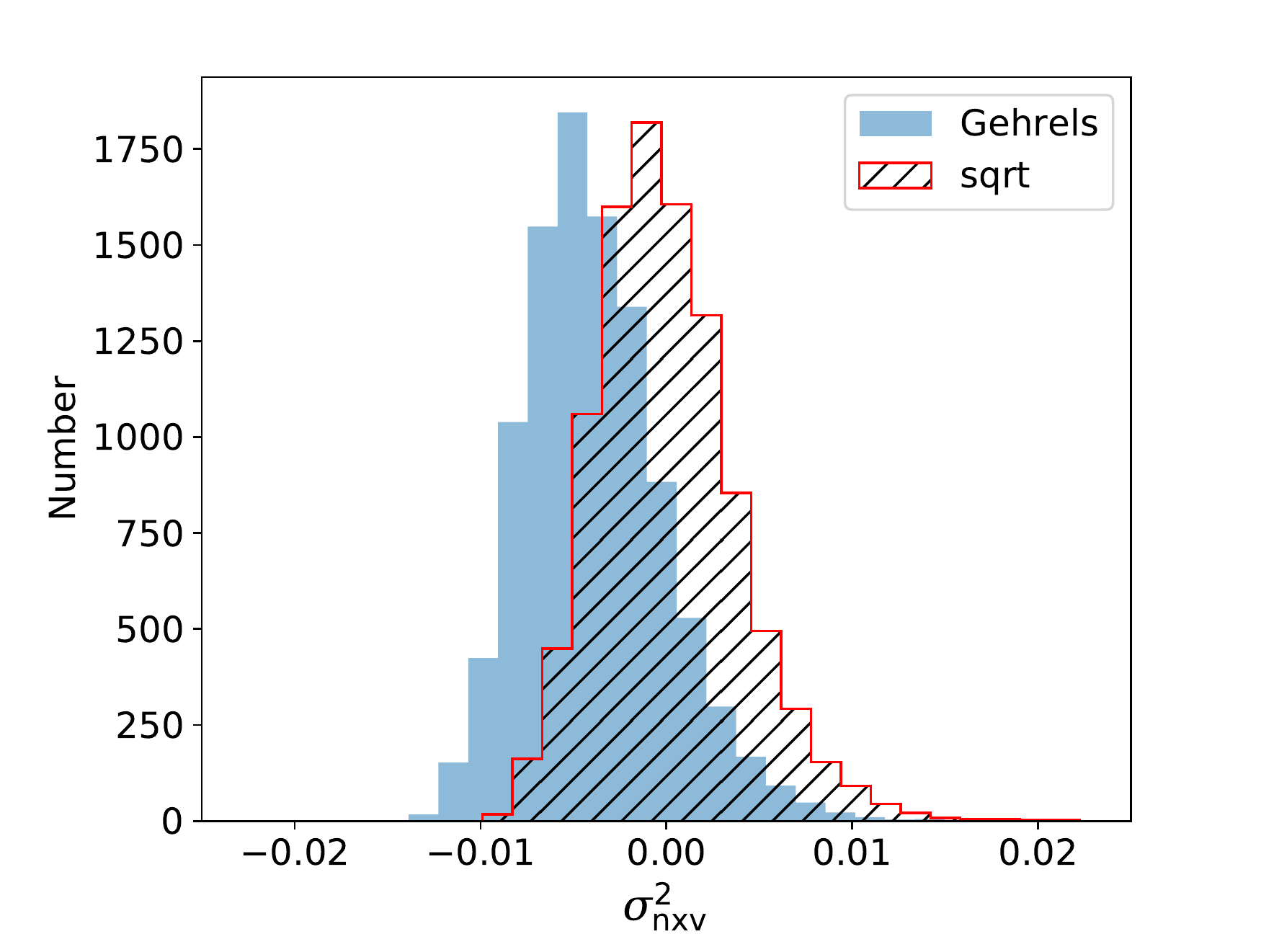}
\caption{Distributions of $\sigma_{\rm nxv}^2$ calculated from simulated light curves of a non-variable source. The blue histogram represents $\sigma_{\rm nxv}^2$ based on the Gehrels error estimation (i.e., Eq.~\ref{eq:geh}). The red hatched histogram represents $\sigma_{\rm nxv}^2$ based on the square root error (i.e., Eq.~\ref{eq:sqerr}).}
\label{fig:error}
\end{figure}

It should be noted that, instead of using $\sigma_{i,\rm err,Geh}$ (i.e., Eq.~\ref{eq:geh}), the computation of $\sigma_{i,\rm err,var}$ is based on the square root of observed counts (i.e., Eq.~\ref{eq:sqerr}) and its corresponding error propagation. This choice has been proven to be a maximum-likelihood estimator for the Gaussian statistic in \citet{Almaini00}; furthermore, \citet{Allevato13} proved that it could also be applied to the low-counts regime.
We also design a test to show the different \nxv behaviors between adopting $\sigma_{i,\rm err,var}$ and $\sigma_{i,\rm err,Geh}$. We simulate 10000 observed light curves of a non-variable source with a mean count rate of about $6\times10^{-5}$~counts~s$^{-1}$ (i.e., about 400 counts in the 7~Ms exposure) with a background level similar to an arbitrary real source. We plot the distributions of $\sigma_{\rm nxv}^2$\ calculated with two kinds of error estimates in Figure~\ref{fig:error}. For a non-variable source, the variable amplitude is 0, so the mean measured \nxv should be close to 0. It is clear that using the Gehrels error ($\sigma_{i,\rm err,Geh}$) yields \nxv values that are systematically smaller than 0. In contrast, \nxv values based on the square root error ($\sigma_{i,\rm err,var}$) are distributed around 0, which means that this estimation is unbiased. 

In addition to the choice of $\sigma_{i,\rm err,var}$, the S/N and total counts also have nonnegligible effects on variability measurement. It has been known that faint sources are more difficult to be classified as being variable \citep[e.g.,][]{Paolillo04,Paolillo17,Lanzuisi14}. \citet{Allevato13} has proven that the uncertainty in \nxv measurement will become larger for sources with lower counts. 
Moreover, irregular sampling patterns can cause additional biases and scatters that could only be quantified through simulations. 

To evaluate the influence of these biases, we perform a test following the procedure below:
\begin{itemize}
 \item[1] We select 30 brightest AGNs (each with $\gsim 2400$ full-band net counts) in our initial source sample to construct a ``bright sample''. These 30 AGNs have very high-quality light curves and can be regarded as sources not influenced by noise.
 \item[2] We randomly choose an AGN in the bright sample, and rescale its full-band light curve such that its average photon flux matches that of an arbitrary fainter AGN (i.e., with $\lsim$ 2400 counts) in the L17 main catalog; note that, in order to show the biased trend more clearly, here we also use faint AGNs with less than 100 total counts. Such a rescaling would not change the variability of the original light curve, so that we could simulate the ``intrinsic'' light curve of a faint source that has the same variability as an AGN in the bright sample.
 \item[3] To simulate the influence of low S/N, we add the Poisson-distributed background (i.e., noise) to the faint ``intrinsic'' light curve, and then extract the ``observed'' counts of each observation. Finally, we obtain a fake light curve of a faint source whose intrinsic variability is the same as that of an AGN in the bright sample.
 \item[4] We repeat steps 2 to 3 1000 times, and compute \nxv of these 1000 simulated faint light curves.
\end{itemize}

\begin{figure}[tp]
\centering
\includegraphics[width=1.1\linewidth]{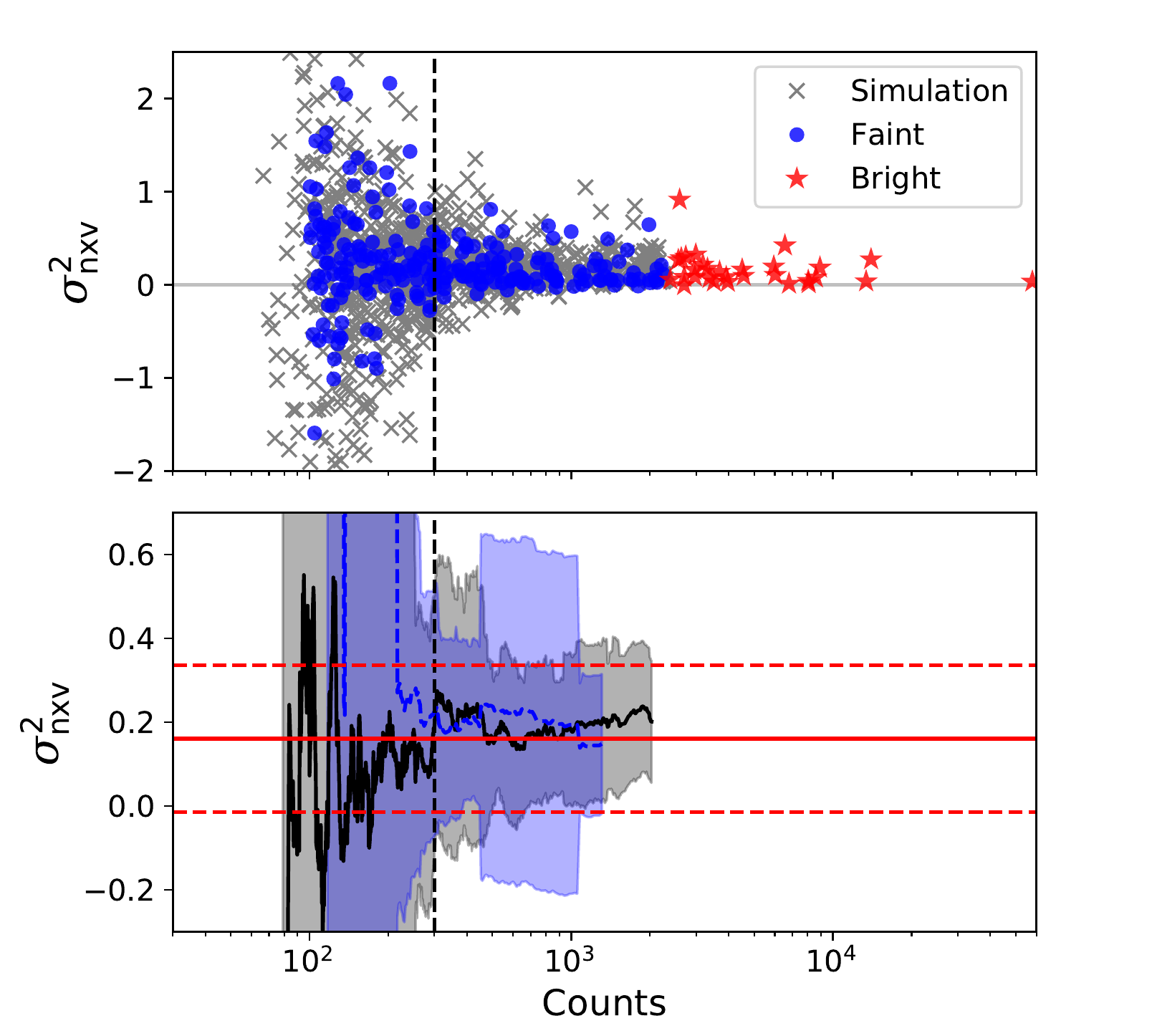}
\caption{Normalized excess variances vs. net counts. (Top) The red stars represent real bright sources that are used in simulation; the blue dots are real faint sources; and the gray crosses denote simulated sources. The vertical black dashed line indicates the minimum counts needed to avoid the influence of noise. Some points with very large/small \nxv values are not shown for clarity. (Bottom) The running averages of the top panel. The blue dashed curve and shaded region stand for the average \nxv and corresponding 1$\sigma$ errors of every 50 neighboring faint sources, while the black curve and shaded region represent those of every 100 simulated sources. The red horizontal solid and dashed lines denote the average \nxv and 1$\sigma$ limits of the bright sample.}
\label{fig:fnxv}
\end{figure}

In the top panel of Figure~\ref{fig:fnxv}, we plot the \nxvt--counts relation of both the real (red and blue symbols) and fake sources (gray symbols). The trend of decreasing scatters of \nxv toward large counts appears apparent and similar for both the real and fake sources. In the bottom panel, we show the running averages and scatters of \nxv for the real faint sources and fake sources. The running bin sizes are 50 for the real faint sources and 100 for the fake sources. The averages and scatters of \nxv are largely similar between the faint and fake samples above $\sim300$ counts and the bright sample, while the scatters of \nxv in the faint and fake samples become unacceptably large below $\sim300$ counts, which can also be inferred from the top panel. 
The similarity in the overall trend of \nxvt--counts and associated scatters between the faint and fake samples suggests that the large \nxv scatters of very faint sources (i.e., having $\lsim300$ counts in this context) originate from low S/N (i.e., being significantly influenced by noise). Fig.~\ref{fig:fnxv} also reflects that, above $\sim300$ counts, there is no significant difference in variability between the bright and faint samples.

It should be noted that, in the above procedure, we find there is a larger fraction of negative \nxv for sources below the 300 counts threshold in the fake sample than that in the real faint sample. This fact can be seen in the bottom panel, where the average \nxv of real data is always positive while that of fake data can sometimes be smaller than zero. This discrepancy should be interpreted as being primarily due to the Eddington bias, i.e., in the low-counts regime, very faint sources with large variability and positive flux fluctuations (thus having positive average \nxvt) are more likely to be detected. Given that we only focus on sources above the 300~counts threshold (see Section~\ref{sec:finsamp}), the Eddington bias would not affect our following analyses.

\subsection{Unbiased sample construction}\label{sec:finsamp}

According to Fig.~\ref{fig:fnxv} and the above arguments, 
it is clear that the influence of noise can be ignored while measuring variability amplitudes of sources with $\gsim 300$ full-band counts. Therefore, we are able to obtain an unbiased sample by applying this counts threshold cut. 

We perform a similar analysis to the hard-band and soft-band light curves, and find that the 300 counts threshold could also be applied to the hard-band light-curve analysis while the soft-band light-curve analysis requires only $\gsim 200$ counts. In most of the remaining analyses, we require our studied light curves to have more than 300 full-band counts, except in Section~\ref{res:bands} where we only use sources with more than 300 hard-band counts and more than 200 soft-band counts. There are still a small number of sources with negative \nxv (13 for the full band), but they will not affect our analysis significantly since we will utilize the so-called ``ensemble excess variance'' \citep{Allevato13} by taking the average \nxv of sources that have similar physical properties.

\begin{figure}[tp]
\includegraphics[width=1.1\linewidth]{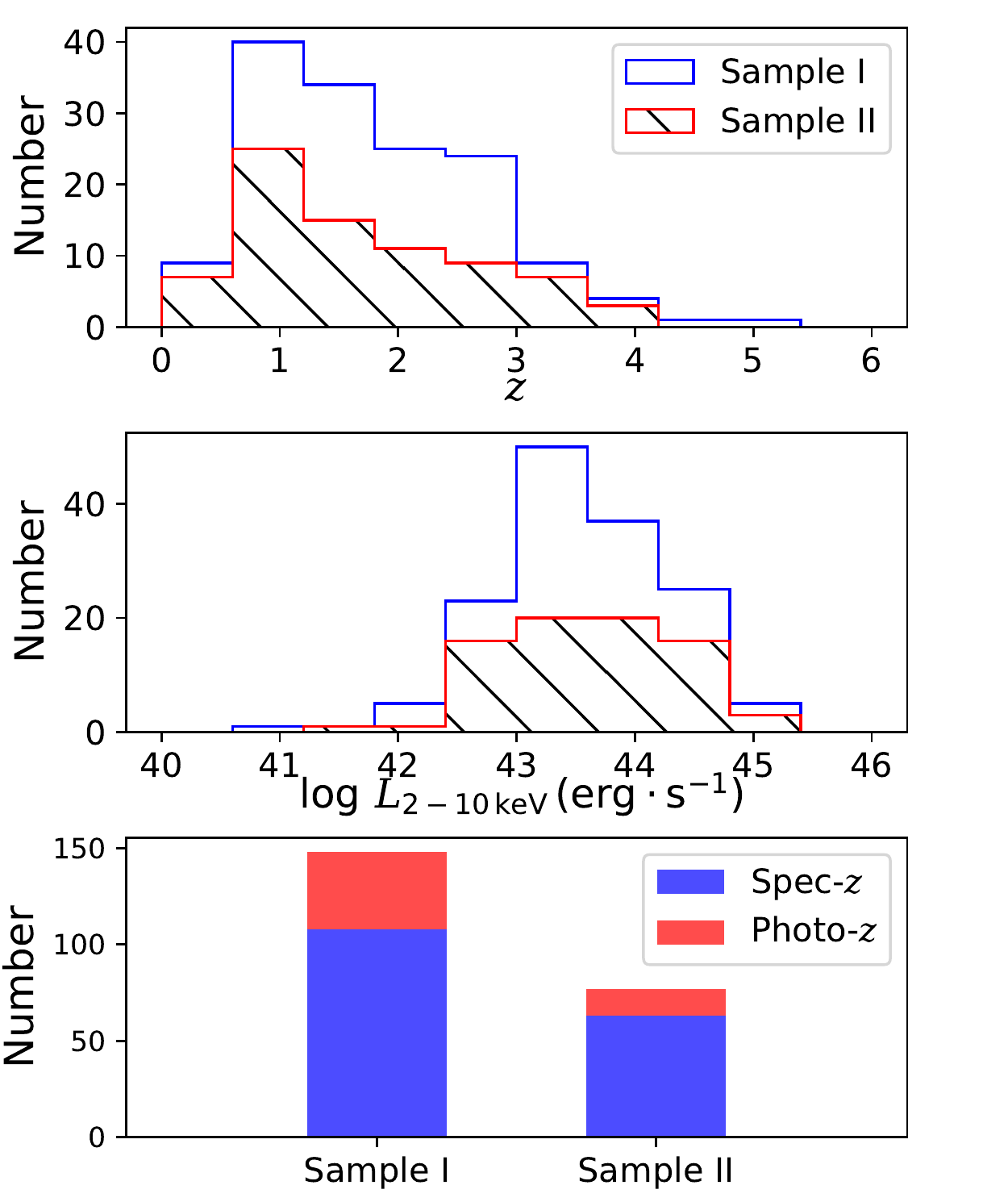}
\caption{(Top) Redshift distributions for the sources in Sample~I (i.e., 148~sources with $\ge 300$ full-band counts) and Sample~II (i.e., 77~sources with $\ge 200$ soft-band counts and $\ge 300$ hard-band counts). (Middle) 2--10 keV luminosity distributions of the two samples. (Bottom) Numbers of spectroscopic and photometric redshifts in the two samples.}
\label{fig:dist}
\end{figure}
Based on the initial sample constructed in Section~\ref{sec:dred}, we find a total of 148 sources whose full-band light curves meet our requirement (i.e., each having $\ge 300$ full-band counts and satisfying the criteria of 1, 3, 4, and 5 in Section~\ref{sec:dred}). These 148 sources make up Sample~\one. Similarly, the numbers of available sources are 110 and 98 for the soft and hard bands respectively, while there are 77 sources that meet the requirements in both the soft (i.e., $\ge 200$ counts) and hard (i.e., $\ge 300$ counts) bands. These 77 sources are marked as Sample~\two. 

\begin{figure*}[tp]
\centering
\includegraphics[width=\textwidth]{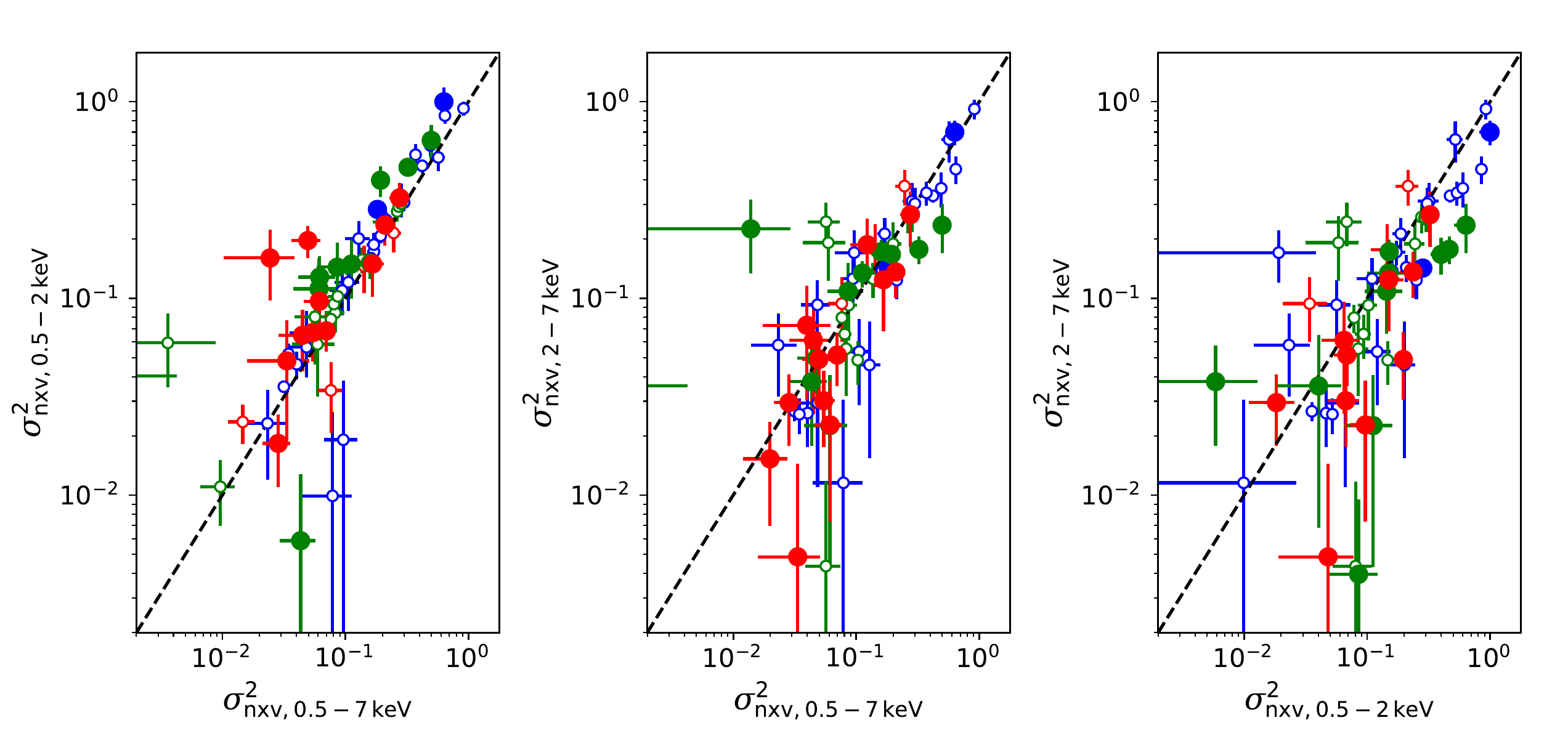}
\caption{Comparison of \nxv from different energy bands using sources in sample \two. (Left): soft-band \nxv vs. full-band \nxvt. (Middle): hard-band \nxv vs. full-band \nxvt. (Right): hard-band \nxv vs. soft-band \nxvt. The filled larger points are sources with $N_{\rm H}$ larger than $10^{22.5}~\rm cm^{-2}$. The colors indicate different source redshifts ($z\le$1: blue; $1<z\le$2: green; $2<z\le$4: red). The dashed line denotes y=x.}
\label{fig:band}
\end{figure*}
Adopting the preferred redshifts in the L17 main catalog (i.e., 
the 51st column, ``ZFINAL''; see Section~4.3 of L17 for the redshift selection criteria) 
and the spectral analysis results of Li et~al. (in prep.), we present the redshift and X-ray luminosity distributions of Sample~I and II sources in Fig.~\ref{fig:dist}. These two samples cover very similar wide ranges of redshift ($0<z\leq 6$) and X-ray luminosity ($\sim10^{41-45}$~erg~s$^{-1}$). 
In Sample~I (Sample~II), 101 (63) sources have spectroscopic-redshift measurements, 
with 83 (50) being secure and 18 (13) being insecure but agreeing well with at least one
of the available photometric-redshift estimates; and the remaining 47 (14) sources have photometric redshifts
as their preferred redshifts, with the 25th, 50th, and 75th percentiles of zphot\_error/(1+zphot)
being 0.018, 0.026, and 0.057 (0.012, 0.019, and 0.026), respectively.
Given the relatively high fractions of spectroscopic redshifts
(101/148=68.2\% for Sample~I and 63/77=81.8\% for Sample~II)
and small uncertainties of photometric redshifts, 
using only (secure) spectroscopic redshifts should not affect
our analysis significantly. Therefore, we choose to use the L17 preferred redshifts, 
which were selected scrutinizingly, in order to maximize our sample sizes.

\section{Results}\label{sec:res}
\subsection{Variability of different energy bands}\label{res:bands}

As mentioned before, we extract our light curves based on observed-frame energy bands, which means that we could discuss variability of different rest-frame energy bands for sources with different redshifts. However, at least for short-term (i.e., $T\lesssim 100~\rm ks$) variability, there is evidence implying that variability amplitudes in various energy bands have a good consistency \citep{Ponti12}. Using the sources in Sample~\two, we compare \nxv measured from light curves in three different bands in Fig.~\ref{fig:band} to check if the consistency remains for long-term variability.

Generally, \nxv in different energy bands are well correlated and the linear slope is close to 1. 
We divide our sources into three subsamples according to their redshifts and mark them with different colors. 
It appears that the correlation behavior of \nxv in different energy bands is largely not influenced at different redshifts, although the subsamples with higher redshifts tend to have relatively larger dispersions.
We also mark the sources with $N_{\rm H}$ larger than $10^{22.5}~\rm cm^{-2}$ using large filled symbols. 
The overall behavior of these obscured sources in Fig.~\ref{fig:band} is quite similar to that of the unobscured sources (see Section~\ref{res:NH} for more details). 

We note that \nxv in the soft band seem to be slightly larger than that in the full band and hard band (see the left and right panels of Fig.~\ref{fig:band}),
which is seen both in Sample II and in the obscured subsample. 
This difference may be explained by the superposition of a soft component varying in flux and/or slope and a constant hard reflection component, which can result in the ``softer when brighter'' behavior \citep[e.g.,][]{Sobolewska09,Gibson12,Serafinelli17}. Additionally, the variability of absorption may be another possible reason, since the soft band is more easily affected by $N_{\rm H}$ variation than the hard band.
However, given that \nxv in the soft band is systematically larger only up to a level of about 10\%--30\%, this difference will not affect materially most of our following analysis except the study of the evolution of variability (see Section~\ref{sec:var2z}).

\subsection{\nxv and $N_{\rm H}$}\label{res:NH}

\begin{figure}[tp]
\includegraphics[width=1.1\linewidth]{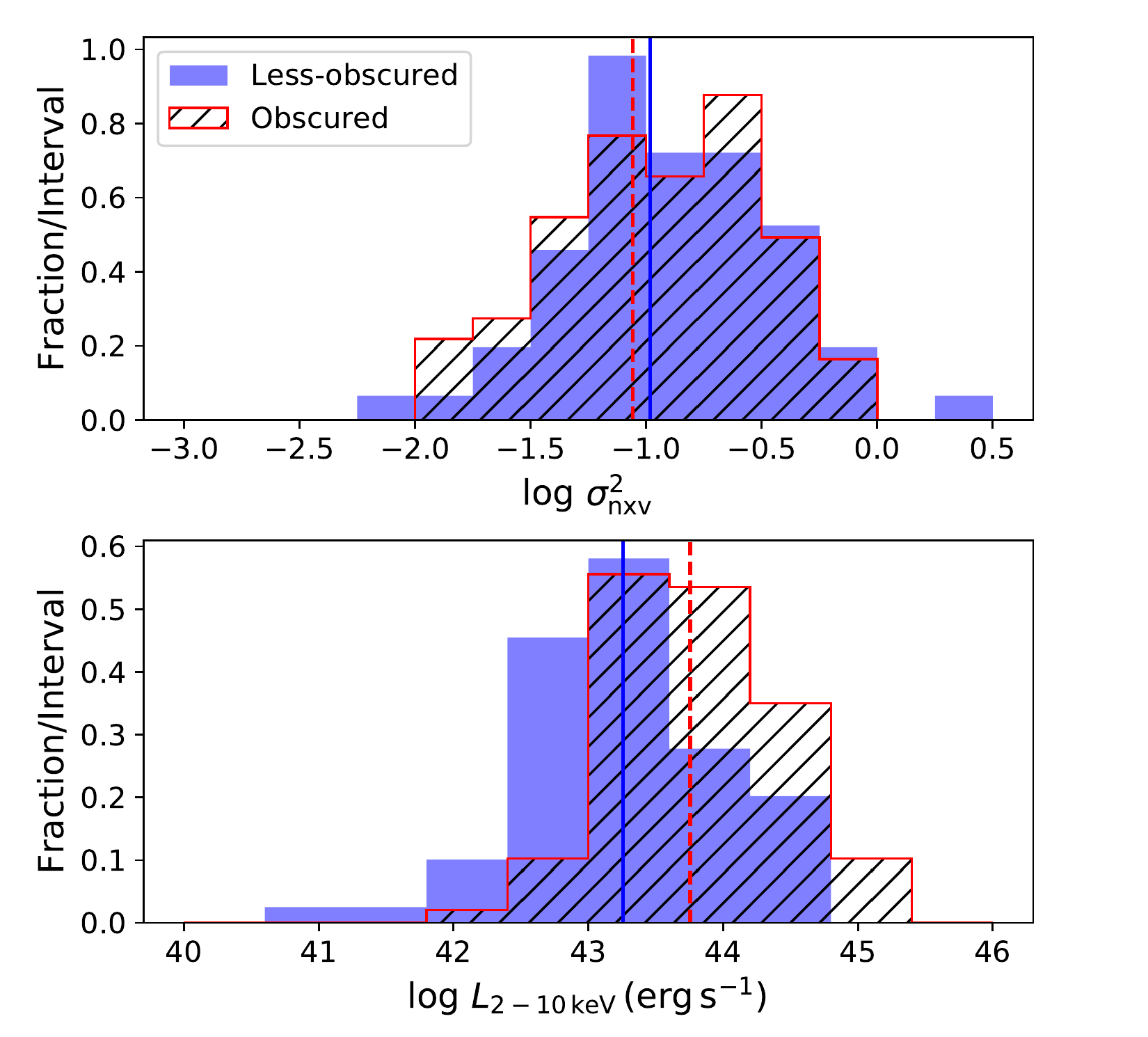}
\caption{Normalized histograms of log~\nxv (top panel) and log~$L_{\rm 2-10keV}$ (bottom panel) for the obscured and less-obscured samples (only sources with \nxv $>0$ are shown). The red hatched histograms and blue histograms are the distributions of obscured and less-obscured sources, respectively. The dashed and solid vertical lines denote the median values of obscured and less-obscured sources, respectively.}
\label{fig:nh}
\end{figure}

Previous studies \citep[e.g.,][]{Paolillo04} found evidence of possible connection between variability and obscuration such that hard obscured AGNs tend to have lower variability. Obscuration might smooth variability and lead to smaller \nxvt. On the other hand, \citet{Yang16} and \citet{Liu17} found some sources with $N_{\rm H}$ variations, which might increase AGN long-term variability. It is not clear how these effects would influence our following analysis. So we divide our sample into two parts: obscured ($N_{\rm H}>10^{23}~\rm cm^{-2}$, 49 sources) and less-obscured ($N_{\rm H}\le 10^{23}~\rm cm^{-2}$, 99 sources) and plot their \nxv distributions in the top panel of Fig.~\ref{fig:nh}.

\begin{figure}[tp]
    \includegraphics[width=1.05\linewidth]{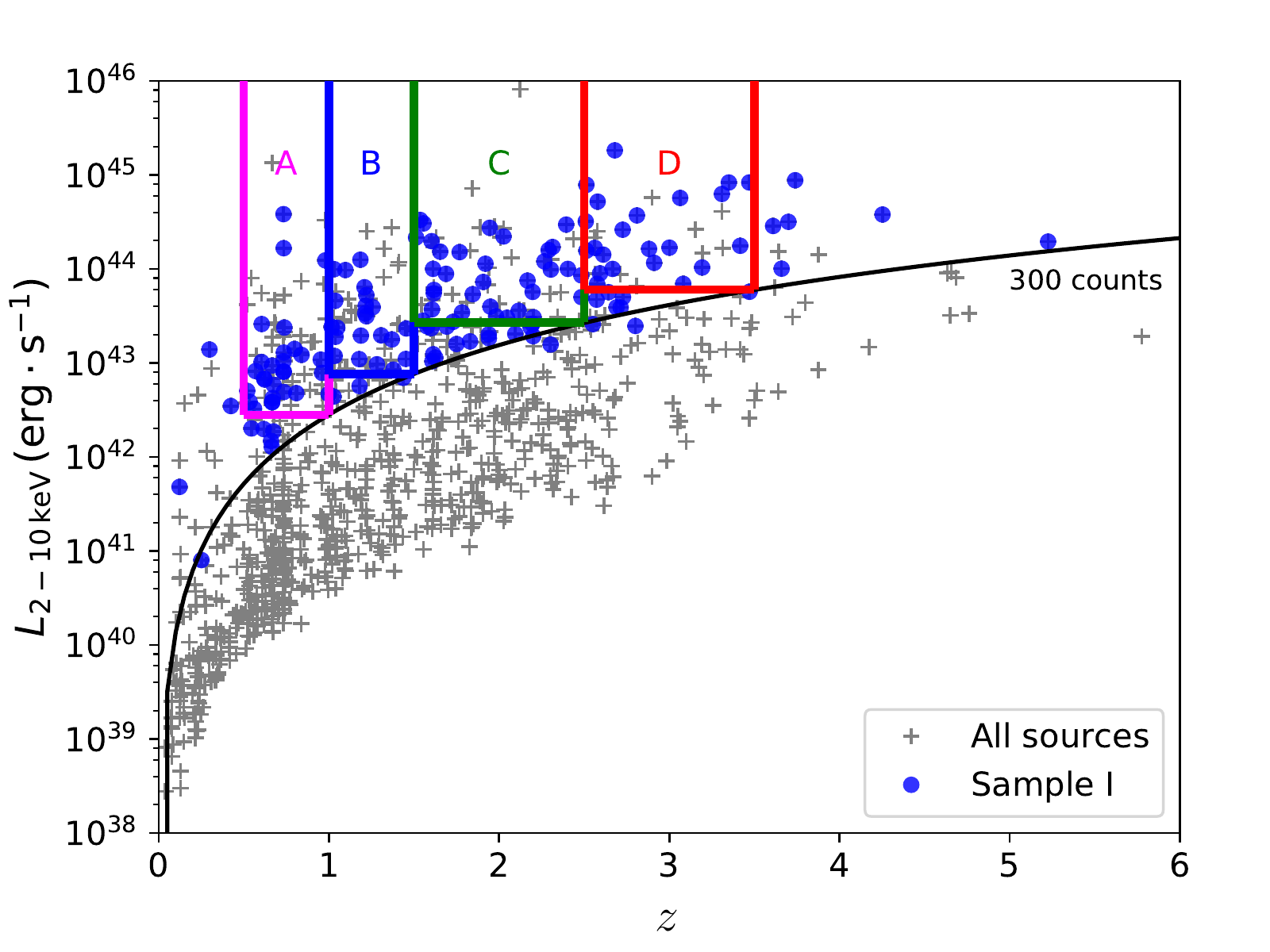}
    \caption{$L_X$ vs. redshifts of our sources. Grey crosses represent all sources in L17. Blue circles represent sources with reliable $\sigma_{\rm nxv}^2$ measurement. Four marked regions denote the subsamples used in Section~\ref{res:NH}, Fig.~\ref{fig:t_nxv}, and Table~\ref{tab:tnxv}. The 300 counts limit used in source selection is marked with the black curve.}
    \label{fig:csample}
\end{figure}

We perform a K-S test to assess the similarity of the two samples and the result indicates that their $\sigma_{\rm nxv}^2$ distributions are quite similar ($P_{\rm reject}\approx 26\%$). However, since obscured sources tend to have larger intrinsic luminosities (in our sample, obscured sources have a mean $L_{\rm 2-10~keV}$ of $\approx 1.3\times10^{44}~\rm erg~s^{-1}$, while less-obscured sources have a mean $L_{\rm 2-10~keV}$ of $\approx 5\times10^{43}~\rm erg~ s^{-1}$; see the bottom panel of Fig.~\ref{fig:nh}), we would expect that they should have smaller $\sigma_{\rm nxv}^2$ based on the known anti-correlation between variability and luminosity. In fact, when we compare the median log~\nxvt, obscured sources do have smaller \nxv values though not significantly ($\Delta \rm log~\sigma_{\rm nxv}^2\sim0.2$~dex). From the results of Section  \ref{res:Lx} and other studies \citep[e.g.,][]{Lanzuisi14}, we find the $L_{\rm X}-\sigma_{\rm nxv}^2$ relation is enough to explain this difference. In order to disentangle the influences of redshift and $L_{\rm X}$ (see Fig.~\ref{fig:csample} for the plot of $L_{\rm X}$ vs. $z$), we also choose 5 complete subsamples (see Table~\ref{tab:nhres}), within which sources have similar redshifts and luminosities, and then perform the Spearman's ranking test to check the correlation between their $N_{\rm H}$ and $\sigma_{\rm nxv}^2$. The results are shown in Table \ref{tab:nhres}, indicating that none of these subsamples shows an evident correlation between $N_{\rm H}$ and variability (i.e., all $P_{\rm reject}$ values are $\ge 8\%$; but note the limited sizes of the subsamples). 

\begin{table}[tp]
    \begin{center}
        \caption{Spearman's ranking test results of $\sigma_{\rm nxv}^2-N_{\rm H}$}\label{tab:nhres}
        \resizebox{\linewidth}{!}{
        \begin{tabular}{cccccc}\hline\hline
            Subsample & Size &$z$ & $L_{\rm 2-10~ keV}$ ($10^{43} \rm erg\ s^{-1}$) & $\rho$ & $P_{\rm reject}$ \\\hline
            1 & 5 & 0.7--1.1 & 3--30 & 0.10 & 0.13\\
            2 & 6 & 1.1--1.5 & 3--30 & $-0.05$ & 0.12\\
            3 & 10& 1.5--2.1 & 3--30 & $-0.06$ & 0.19\\
            4 & 10& 2.1--2.8 & 3--30 & 0.02 & 0.08\\
            5 & 10& 0.7--1.1 & 0.8--3& 0.37 & 0.79\\
            \hline
        \end{tabular}
        }
    \end{center}
\end{table}

Based on the above results of the K-S test and Spearman's ranking tests, we conclude that the subsequent analysis of $L_{\rm X}-\sigma_{\rm nxv}^2$ does not suffer from the bias caused by obscuration.

\subsection{\nxv vs. $L_{\rm x}$}\label{res:Lx}

It has long been known that X-ray variability amplitude is well anti-correlated with luminosity \citep[e.g.,][]{Nandra97,Paolillo04,Paolillo17,Papadakis08,Gonzalez11,Ponti12,Lanzuisi14,Yang16}. This trend can be a result of the dependence of AGN PSD on the black hole mass and accretion rate. 

\begin{figure}[tp]
\includegraphics[width=1.1\linewidth]{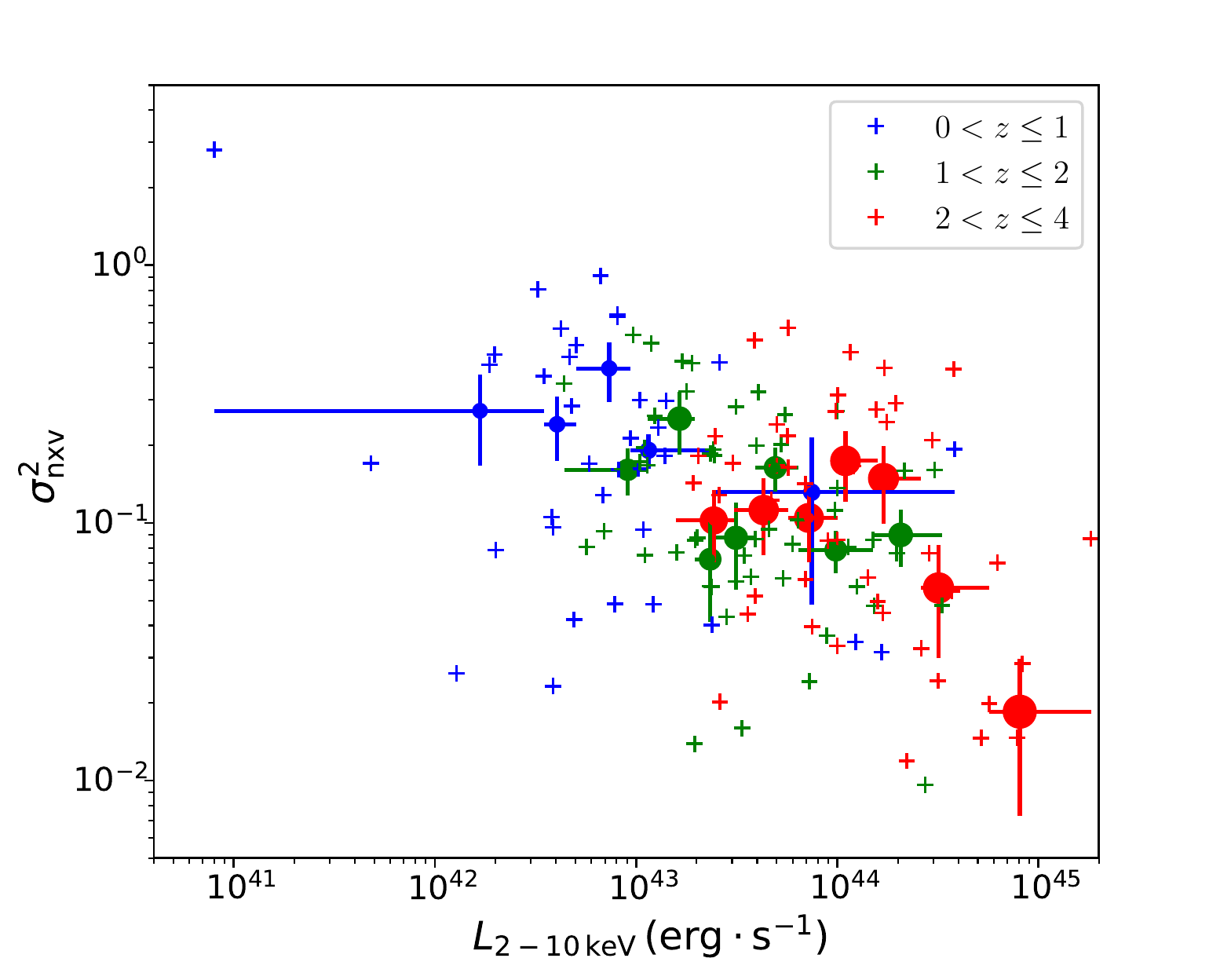}
\caption{\nxv vs. X-ray luminosity for the sources in Sample~I. Colors indicate the redshift ranges. Filled circles are binned results. The positions of binned points are determined by the medians of $L_{\rm 2-10~keV}$ and mean values of \nxvt, and the sizes represent their average redshifts. Every $L_{\rm 2-10~keV}$ bin contains 8 sources from a same subsample (i.e., their redshifts are close). Y-axis error bars are derived from standard errors and X-axis error bars show the luminosity ranges.}
\label{fig:lx}
\end{figure}

In Fig.~\ref{fig:lx} we display the \nxvt--$L_{\rm 2-10~keV}$ relation of all sources in sample~I as defined in Section~\ref{sec:finsamp}. A decreasing trend is revealed, but the trend may be not as apparent if we only look at one subsample with a certain range of redshifts because of the large scatter and relatively narrow $L_{\rm 2-10~keV}$ range. Therefore we bin our data and plot them in Fig.~\ref{fig:lx}. We only bin sources in a same subsample that have similar redshifts, because \nxv for different redshifts stands for the variability of different rest-frame timescales. Each binned data point represents an average \nxv of 8 sources with close $L_{\rm 2-10~keV}$ values. The bin size is chosen to balance the luminosity range within each bin and the requirement of reliable average \nxv calculation. The error bars denote standard errors and luminosity ranges. The symbol sizes denote the average redshifts of the bins.

After binning, we see a clear anti-correlation between \nxv and $L_{\rm 2-10~keV}$ for the whole sample, which is also manifested by the Spearman's ranking test results based on individual sources, although the trend is not significant for either the low-redshift or high-redshift subsample (for all sources, $\rho=-0.31$, $P_{\rm reject}=10^{-4}$; for $z<1.5$ sources, $\rho=-0.17$, $P_{\rm reject}=0.2$; and for $z\geq 1.5$ sources, $\rho=-0.17$, $P_{\rm reject}=0.1$). It should be noted that the reason we perform tests to $z<1.5$ and $z\geq 1.5$ sources instead of the three subsamples we use in binning, is that the luminosity range of any of the three subsamples is narrow.

The decreasing trend of \nxv toward large $L_{\rm 2-10~keV}$ might be due to two reasons: time dilution (due to redshift) and PSD shape. As we know, \nxv is the integral of PSD:
\begin{equation}
\sigma_{\rm nxv}^2 = \int_{\frac{1}{T_{\mathrm{rest}}}}^{\frac{1}{2\Delta t_{\mathrm{rest}}}} \mathrm{PSD} (\nu)\mathrm{d}\nu, \label{eq:int}
\end{equation}
where $T_{\rm rest}$ and $\Delta t_{\mathrm{rest}}$ are the length of light curve and the bin size\footnote{Due to irregular sampling, the bin size is not a constant, therefore we set $\Delta t_{\rm obs}\approx 80 \rm ks$ in our analysis since it is a typical length of the observations.} in the rest frame, respectively; and the PSD is often assumed to be a single or broken power law. Although our sources have similar observational exposures and sampling patterns, their large redshift range makes a big difference to their rest-frame timescales. Therefore, for high-redshift AGNs, their integrating intervals in Eq.\ref{eq:int} will shift to higher-frequency ranges because their light curves are shorter in the rest frame. If the AGN PSD follows a uniform power law $\mathrm{PSD}(\nu)\propto \nu^{-\beta}$, sources with higher redshifts are supposed to have smaller \nxv in our measurement if $\beta>1$, because Eq.\ref{eq:int} would become
\begin{equation}
    \sigma_{\rm nxv}^2 = (1+z)^{-\beta+1}\int_{\frac{1}{T_{\mathrm{obs}}}}^{\frac{1}{2\Delta t_{\mathrm{obs}}}} \mathrm{PSD} (\nu)\mathrm{d}\nu.\label{eq:int2} 
\end{equation}

From Eq.~\ref{eq:int2}, the influence of redshift uncertainties can also be estimated. 
As demonstrated in Section~\ref{sec:finsamp}, the uncertainties of our adopted photometric 
redshifts are relatively small, the majority of which have values of zphot\_error/(1+zphot) less than
a few percent. Even when $\beta=1.5$, the resulting deviation is only about 20\% 
considering the photometric redshifts that have the largest uncertainties. 
Since we use average \nxv in the subsequent fitting, this influence will be further reduced.

Another influence comes from PSD shape.\footnote{We note that the PSD models 
discussed in both this subsection and Section~\ref{sec:psdmod} are purely empirical based
on local AGN studies, which could be the observational manifestation of the various underlying
physical processes, such as the superposition of many randomly flaring subunits \citep{Green93, Nandra97} or a relation between the luminosity and the size of a single varying region \citep{Almaini00}.
However, the variability analyses presented here would not be able to constrain those theoretical considerations.} 
As mentioned before, the AGN PSD can be well represented by a broken power law. Previous studies \citep[e.g.,][]{McHardy06,Gonzalez12,Ponti12} pointed out that the high-frequency break depends on black hole mass and Eddington ratio $\lambda_{\rm Edd}$, which could be expressed as $\nu_{\rm hb}\propto M_{\rm BH}^{-1}\lambda_{\mathrm{Edd}}^{\gamma}$, where the value of $\gamma$ is still controversial. In addition, the normalization of PSD is found to be roughly inversely proportional to $\nu_{\rm hb}$ \citep{Papadakis04}. In Section~\ref{sec:psdmod}, these results will be introduced. Consequently, assuming PSD ($\nu$)=$A (\nu/\nu_{\rm hb})^{-2}$ when $\nu>\nu_{\rm hb}$, we would derive $\sigma_{\rm nxv}^2\sim A\nu_{\rm hb}^2 T_{\rm rest}$, which could also contribute to the anti-correlation between \nxv and $L_{\rm 2-10~keV}$. However, the lengths of our light curves are over 16 years, which means $1/T\sim 2\times10^{-9} \ll \nu_{\rm hb}$. Moreover, since most of our observations lasted for $10^{4}-10^{5}\rm s$, the corresponding upper bound of integral $1/2\Delta t$ in Eq.\ref{eq:int2} is close to $\nu_{\rm hb}$ for supermassive black holes \citep[e.g.,][]{McHardy06,Gonzalez12}. This means that our $\sigma_{\rm nxv}^2$ are more likely to be dominated by the low-frequency part of PSD. Some studies assumed a power law PSD with an index of 1 when $\nu<\nu_{\rm hb}$. But the exact form of the low-frequency AGN X-ray PSD still needs to be explored with the help of longterm monitoring data.

Therefore, we take into account the bin size and the power law indexes of different parts of PSD to fit our \nxvt-$L_{\rm 2-10~keV}$ results, and try to figure out how these parameters affect the observed anti-correlation trend. Furthermore, the bias caused by irregular sampling needs to be assessed with the use of light curve simulations assuming a certain type of AGN PSD.
It should be noted that similar PSD analyses could also be found in \citet{Paolillo17}, where they tried to study the accretion history of SMBHs while we aim to constrain the exact form of AGN PSD.

\subsection{PSD modelling}\label{sec:psdmod}

\begin{figure*}[tp]
\includegraphics[width=\textwidth]{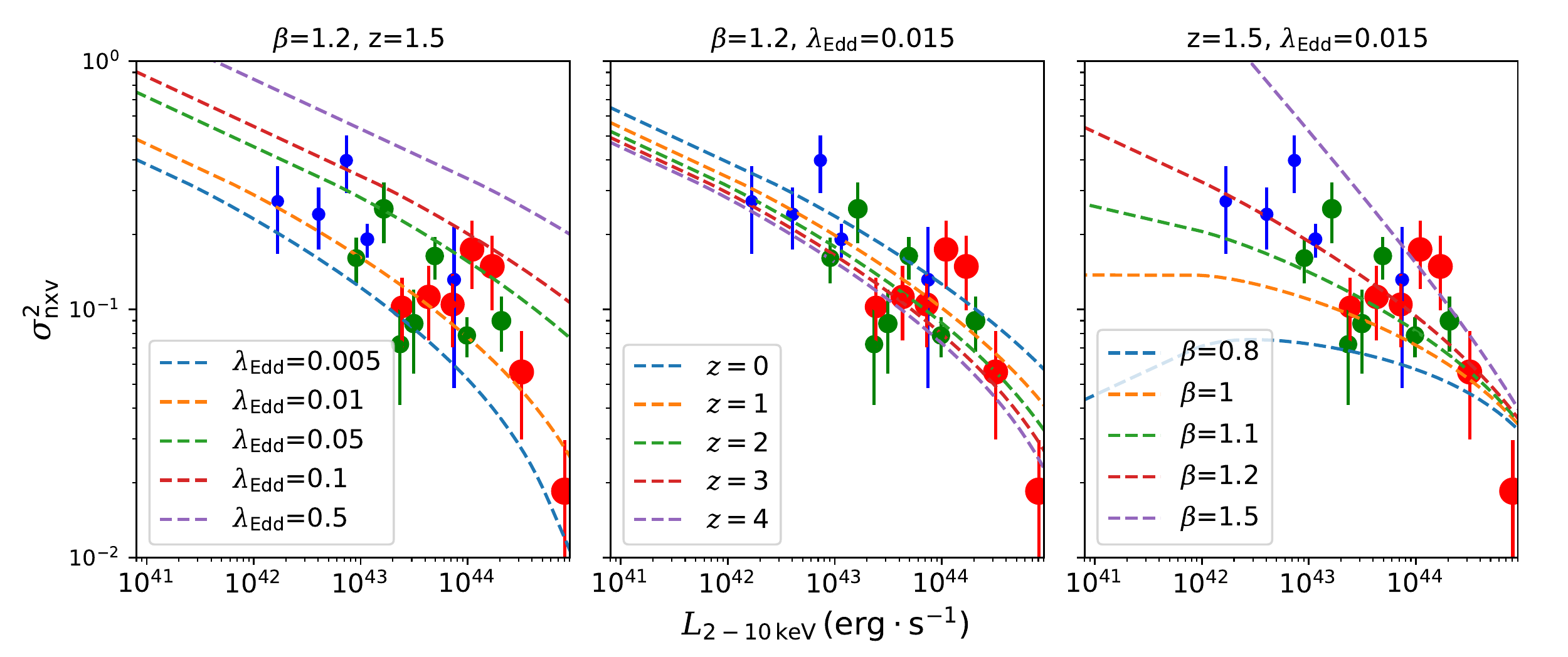}
\caption{Comparison between observed and model $\sigma_{\rm nxv}^2-L_{\rm 2-10~keV}$ relations using the \citet{McHardy06} $\nu_{\rm hb}$ computation. Data points are the same as those in Fig.~\ref{fig:lx}. (Left) The influence of Eddington ratio. The change in accretion rate leads to a shift along the $L_{\rm 2-10~keV}$ axis. Other parameters are annotated above the plot. (Middle) The influence of redshift. Higher-redshift sources show slightly smaller $\sigma_{\rm nxv}^2$ values. (Right) The influence of $\beta$. Larger $\beta$ values lead to larger $\sigma_{\rm nxv}^2$ values. The change in $\beta$ also makes a difference to the shape of the relation.} 
\label{fig:lxmod}
\end{figure*}

\begin{figure*}[tp]
    \begin{center}
    \includegraphics[width=\linewidth]{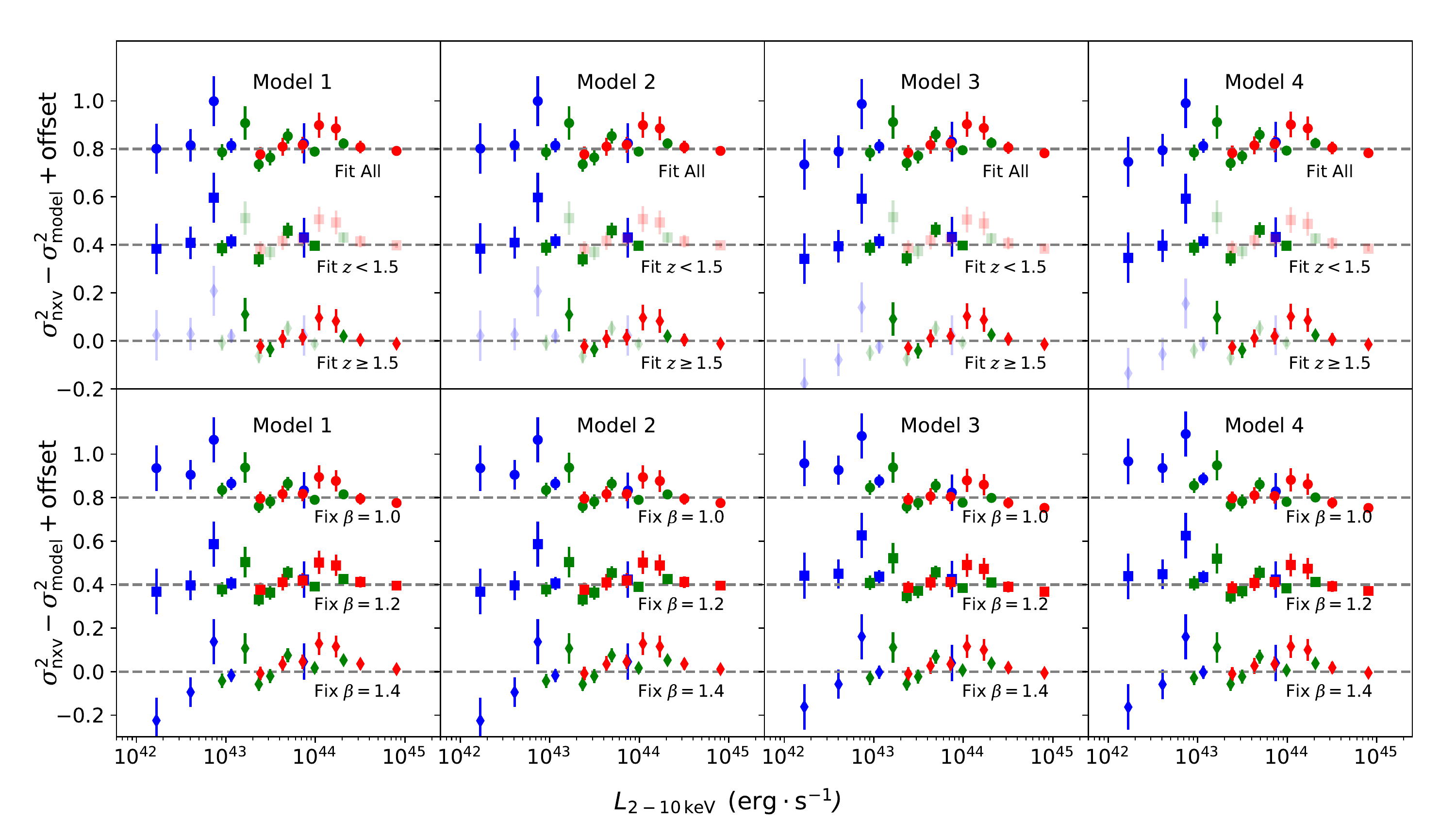}
    \caption{Fitting results of the $L_{\rm 2-10\rm keV}-\sigma_{\rm nxv}^2$ relation, shown by residual plots. (Top) Fitting results of 4 models. Grey dashed lines stand for residual = 0. For points with different redshift ranges, we mark with different colors. The results are offset for clarity. Points used in the fitting are highlighted and labeled. We can see all models can fit the data well with proper parameters. 
     (Bottom) Fitting results of 4 models with $\beta$ fixed. Circles stand for results for $\beta =1$, while squares and diamonds are for $\beta=1.2$ and $\beta=1.4$, respectively. All points are used in this kind of fitting. The choice of $\beta$ has a significant influence on fitting the low-redshift low-luminosity sources.}\label{fig:fit}
    \end{center}
\end{figure*}

Previous studies (\citealp[e.g.,][]{Gonzalez11}; also see, e.g., Fig.~1 of \citealt{Zhu16} for an illustration) suggest that the AGN PSD can be expressed as
\begin{equation}
    \mathrm{PSD} (\nu)=\left\{
        \begin{array}{ll}
            A (\nu/\nu_{\rm hb})^{-\alpha} & (\nu>\nu_{\rm hb}) \\
            A (\nu/\nu_{\rm hb})^{-\beta} & (\nu_{\rm hb}\ge\nu>\nu_{\rm lb}), \\
            A (\nu_{\rm lb}/\nu_{\rm hb})^{-\beta} & (\nu\le\nu_{\rm lb})
        \end{array}
        \right.\label{eq:psd}
\end{equation}
where the high-frequency slope $\alpha$ is close to 2 while the low-frequency slope $\beta$ is found to be about 1 in some bright sources \citep[e.g.,][]{Uttley05,Breedt09}. In \citet{Papadakis04}, it was found that $C_1=A\nu_{\rm hb}$ is roughly a constant of 0.017. Furthermore, although the ratio $\nu_{\rm lb}/\nu_{\rm hb}$ is about 0.1 in Galactic BHBs, a study on Ark~564 reported a ratio of about $10^{-4}$ \citep{McHardy07}. But as shown in Fig.~\ref{fig:lx}, there is no sign of a second PSD break, which suggests that the very low frequency part of PSD does not play an important role in the relation. Therefore we only consider the high-frequency break and PSD normalization in the following analysis.

\begin{table}[tp]
    \begin{center}
    \caption{$L_{\rm 2-10~\rm keV}-\sigma_{\rm nxv}^2$ Fitting Results}\label{tab:fit}\begin{tabular}{cccccc}\hline\hline
        Sample& Model &Typical log $\lambda_{\mathrm{Edd}}$ & $\beta$ & $\chi^2_{\nu}$ & d.o.f\\ \hline
        All & 1 & $-1.82^{+0.12}_{-0.11}$ & $1.16^{+0.05}_{-0.05}$ & 1.4 & 17\\
        All & 2 & $-3.35^{+0.28}_{-0.25}$ & $1.16^{+0.05}_{-0.05}$ & 1.4 & 17\\
        All & 3 & $0.6^{+0.5}_{-0.4}$ & $1.31^{+0.04}_{-0.04} $ & 1.6 & 17 \\
        All & 4 & $0.07^{+0.22}_{-0.19} $ & $1.30^{+0.04}_{-0.04}$ & 1.5 & 17\\
        $z<1.5$ & 1 & $-1.94^{+0.21}_{-0.18}$ & $1.20^{+0.07}_{-0.07}$ & 1.6 & 7\\
        $z<1.5$ & 2 & $-3.6^{+0.5}_{-0.4}$ & $1.20^{+0.07}_{-0.07}$ & 1.6 & 7\\
        $z<1.5$ & 3 & $0.7^{+1.0}_{-0.6}$ & $1.31^{+0.06}_{-0.06}$ & 1.6 & 7\\
        $z<1.5$ & 4 & $0.13^{+0.37}_{-0.30}$ & $1.31^{+0.06}_{-0.06}$ & 1.6 & 7\\
        $ z\ge 1.5$ & 1 & $-1.77^{+0.16}_{-0.15}$ & $1.14^{+0.10}_{-0.10}$ & 1.5 & 8\\ 
        $ z\ge 1.5$ & 2 & $-3.22^{+0.37}_{-0.34}$ & $1.14^{+0.10}_{-0.10}$ & 1.5 & 8\\ 
        $ z\ge 1.5$ & 3 & $1.2^{+1.7}_{-1.0}$ & $1.37^{+0.07}_{-0.09}$ & 1.8 & 8\\ 
        $ z\ge 1.5$ & 4 & $0.2^{+0.5}_{-0.4}$ & $1.34^{+0.10}_{-0.10}$ & 1.8 & 8\\ 
        \hline
        All & 1 & $-1.51^{+0.11}_{-0.11}$ & $ 1 (f)*$ & 2.1 & 18\\
        All & 2 & $-2.62^{+0.27}_{-0.26}$ & $ 1 (f)$ & 2.1 & 18\\
        All & 3 & $-0.85^{+0.05}_{-0.05}$ & $ 1 (f)$ & 3.1 & 18\\
        All & 4 & $-0.80^{+0.04}_{-0.04}$ & $ 1 (f)$ & 3.3 & 18\\
        All & 1 & $-1.89^{+0.06}_{-0.07}$ & $ 1.2 (f)$ & 1.4 & 18\\
        All & 2 & $-3.50^{+0.14}_{-0.15}$ & $ 1.2 (f)$ & 1.4 & 18\\
        All & 3 & $-0.22^{+0.06}_{-0.05}$ & $ 1.2 (f)$ & 1.8 & 18\\
        All & 4 & $-0.32^{+0.04}_{-0.04}$ & $ 1.2 (f)$ & 1.7 & 18\\
        All & 1 & $-2.29^{+0.04}_{-0.05}$ & $ 1.4 (f)$ & 2.6 & 18\\
        All & 2 & $-4.39^{+0.10}_{-0.11}$ & $ 1.4 (f)$ & 2.6 & 18\\
        All & 3 & $2.15^{+0.12}_{-0.11}$ & $ 1.4 (f)$ & 1.8 & 18\\
        All & 4 & $0.64^{+0.05}_{-0.05}$ & $ 1.4 (f)$ & 1.8 & 18\\
        \hline
    \end{tabular}
    \end{center}
    {\sc Note.} --\\
    * Label (f) means that the parameter $\beta$ is fixed in the fitting.
\end{table}

We test four models, labeled below as Model 1 to 4, that link PSD to black hole mass and Eddington ratio $\lambda_{\mathrm{Edd}}$:
\\
\begin{itemize}
    \item[1.] We use the $\nu_{\rm hb}$ computation given by \citet{McHardy06},
        $$
        \nu_{\rm hb}=0.003\lambda_{\mathrm{Edd}} (M_{\rm BH}/10^6M_{\odot})^{-1},
        $$
        assuming the PSD amplitude $\nu_{\rm hb}\times \mathrm{PSD} (\nu_{\rm hb})=0.017$ as suggested by \citet{Papadakis04}.
    \item[2.] We adopt the same PSD amplitude as in Model~1, but use the break frequency computed according to \citet[also see \citealt{Pan15}]{Gonzalez12}:
        $$
        \nu_{\rm hb}=0.001\lambda_{\mathrm{Edd}}^{0.24} (M_{\rm BH}/10^6M_{\odot})^{-1}.
        $$
    \item[3.] We use the same break frequency as in Model~1, but adopt the PSD amplitude that depends on Eddington ratio as suggested by \citet{Ponti12}:
        $$
        \nu_{\rm hb}\times \mathrm{PSD} (\nu_{\rm hb}) = 0.003\lambda_{\mathrm{Edd}}^{-0.8}.
        $$
    \item[4.] We adopt the break frequency in \citet{Gonzalez12} and the PSD amplitude in \citet{Ponti12}.
\end{itemize}

\begin{figure*}[tp]
    \begin{center}
    \includegraphics[width=\linewidth]{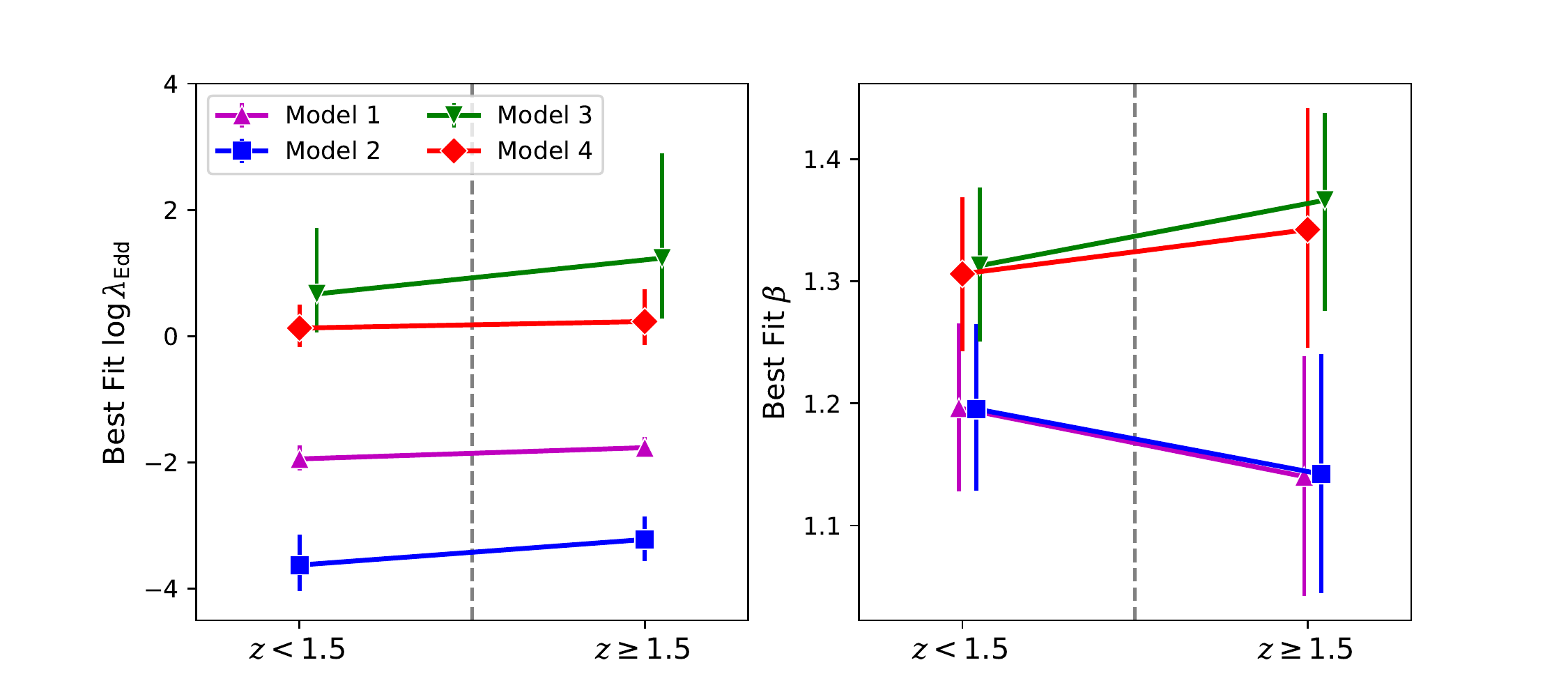}
    \caption{Comparison of best-fit log$~\lambda_{\mathrm{Edd}}$ and $\beta$ for the low-redshift and high-redshift subsamples with different models. Though not significant, all models infer a larger $\lambda_{\mathrm{Edd}}$ at high redshifts. Some points are shifted for clarity.}
    \label{fig:mccomp}
    \end{center}
\end{figure*}
We then use the empirical relation between bolometric correction $k_{\rm bol}$ and $\lambda_{\rm Edd}$, which is computed by studying spectral energy distributions \citep{Lusso10}, to calculate $L_{\rm bol}$ and $M_{\rm BH}$ from $L_{2-10~\rm keV}$ for a given $\lambda_{\rm Edd}$.
Based on these assumptions and Eq.\ref{eq:int}, we are able to connect PSD and $L_{2-10~\rm keV}$, and derive model $L_{\rm 2-10~\rm keV}-\sigma_{\rm nxv}^2$ relations and compare with real data.

In Fig~\ref{fig:lxmod}, we show how different parameters affect the $L_{2-10~\rm keV}-\sigma_{\rm nxv}^2$ relation using Model 1. Generally, the observed $L_{\rm 2-10~\rm keV}-\sigma_{\rm nxv}^2$ relation could be explained by this PSD model with proper parameters. Particularly, we may see that the relation is sensitive to $\lambda_{\mathrm{Edd}}$ and $\beta$. In contrast, if there was a universal PSD, sources with different redshifts would have close \nxv values in our observations, which is consistent with our expectation in Section~\ref{res:Lx} that the uncertainties of redshifts would not affect \nxv significantly. 

Based on these PSD models, we use the \textrm{emcee} code \citep{Foreman13}, which is based on the maximum-likelihood Markov Chain Monte Carlo (MCMC) method, to fit our observed relations and show the results in Fig.~\ref{fig:fit}, Fig.~\ref{fig:mccomp}, and Table \ref{tab:fit}. The best-fit values and their error bars are the median, 16\%, and 84\% percentiles of parameter distributions in the MCMC simulation, respectively. 

With each model, we could find a set of best-fit parameters to fit the data well. When $\beta$ is not fixed in the fitting (see the top row of Fig.~\ref{fig:fit} and the top part of Table~\ref{tab:fit}), we obtain a typical Eddington ratio of about 0.015 and $\beta\approx 1.2$ with Model 1. Using Model 2 leads to a much smaller Eddington ratio of less than $10^{-3}$, which appears a bit small for black hole growth, while $\beta$ is consistent with Model 1. The fitting results of Models 3 and 4 seem to be unrealistic, given that all best-fit $\lambda_{\mathrm{Edd}}$ values are close to or larger than 1, which implies that AGNs are in the super-Eddington accreting state all the time. However, despite of the implausibly large $\lambda_{\mathrm{Edd}}$ values, these two models also suggest a low-frequency PSD index of $\beta\approx 1.3$, very similar to the results for Models 1 and 2.

Comparing the fitting results for $z<1.5$ and $z\geq 1.5$ subsamples in Fig.~\ref{fig:mccomp}, we find a weak tendency that $\lambda_{\mathrm{Edd}}$ is larger in the high-redshift subsample for all models, but the difference is too small compared with the uncertainties. Same as $\lambda_{\mathrm{Edd}}$, the variation of $\beta$ is not apparent as well. We cannot draw a reliable conclusion about whether there is indeed an evolution only with these model-fitting results. This problem will be discussed further in Section~\ref{sec:var2z}.

We notice that the results above are not consistent with \citet{Paolillo17}. It is probably because we do not fix the low-frequency index $\beta$ and/or because our results are based on only one long timescale and thus less sensitive to the break position and more to the PSD normalization.

So we also try to fit the data with $\beta$ fixed to see if a larger $\beta$ is necessary. The results are shown in the bottom row of Fig.~\ref{fig:fit} and the bottom part of Table~\ref{tab:fit}. Apparently, for low-luminosity and low-redshift sources that have longest rest-frame light curves and highest break frequencies (i.e., being most sensitive to the low-frequency part of PSD), model $\sigma_{\rm model}^2$ are too small when $\beta$ is fixed to 1; when $\beta$ is fixed to 1.4, on the contrary, model $\sigma^2_{\rm model}$ are too large. The only well-fitted situation is when $\beta$ is fixed to 1.2, which is very close to the results inferred from the top part of Table~\ref{tab:fit}. We also compare the results with those in \citet{Paolillo17}. Within the uncertainties, our results when $\beta=1$ are in agreement with their results. 

It should be noted that irregular sampling and red-noise leakage \citep[e.g.,][and references therein]{Allevato13,Zhu16} may introduce a bias to the estimation, making
\begin{equation}
\sigma_{\rm nxv,corr}^2 =b \int_{\frac{1}{T}}^{\frac{1}{2\Delta t}} PSD (\nu)\mathrm{d}\nu.\label{eq:bint}
\end{equation}
The bias factor $b$ could only be obtained through simulation especially when both intervals between observations and observation times are irregular and make the choice of $\Delta t$ ambiguous. Therefore, according to the fitting results, we use the light curve simulating code in \citet{Zhu16} to generate 2000 light curves assuming a PSD model whose $\beta=1.2$, $\alpha=2$, and $\nu_{\rm hb}=5\times 10^{-4}~\rm Hz$. We calculate $b$ of these simulated light curves and find that $b\approx 1$ for all redshifts. This result indicates that our \nxv estimation and thus $\sigma_{\rm nxv}^2-L_{\rm 2-10~keV}$ fitting are not subject to the bias caused by irregular sampling and red-noise leakage.

\subsection{Variability of different timescales}\label{sec:tnxv}
Based on Eq. \ref{eq:int} and Eq. \ref{eq:psd}, we may also do a simple estimation of $\beta$ using the $\sigma_{\rm nxv}^2-T$ relation. Assuming $\beta\neq1$ and the high-frequency break $\nu_{\rm hb}=1/t_{\rm hb}$ is between $\nu_1=1/T$ and $\nu_2=1/(2\Delta t)$, we will have
\begin{eqnarray}
	\sigma_{\rm nxv}^2 &=& \int_{\frac{1}{T}}^{\frac{1}{2\Delta t}} PSD(\nu)\mathrm{d}\nu \nonumber\\
	&=& \frac{A}{\beta-1}\frac{T^{\beta-1}}{t_{\rm hb}^{\beta}}-C\quad(\beta\neq1) \label{eq:nxvt1}\\
	C&=& \frac{A}{(\beta-1) t_{\rm hb}}+\frac{A}{(\alpha-1) t_{\rm hb}}[(\frac{2\Delta t}{t_{\rm hb}})^{\alpha-1}-1]\label{eq:nxvt2}.
\end{eqnarray}
For $n_{\rm src}$ sources with similar $T$ and $\Delta t$ but different $A$ and $C$, we will have average \nxv as follows:
\begin{equation}
\langle\sigma_{\rm nxv}^2\rangle = \langle\frac{A}{(\beta-1)t_{\rm hb}^{\beta}}\rangle T^{\beta-1}-\langle C\rangle\label{eq:nxvt3}.
\end{equation}
If $t_{\rm hb}\leq2\Delta t$, Eq. \ref{eq:nxvt2} will become
\begin{equation}
C=\frac{A}{\beta-1}\frac{(2\Delta t)^{\beta-1}}{t_{\rm hb}^{\beta}}\label{eq:nxvt4}.
\end{equation} 
Taking the redshifts into account, we write down the equation in the observed frame as
\begin{equation}
    \langle\sigma_{\rm nxv,obs}^2\rangle = (1+z)^{-\beta+1}(\langle\frac{A}{(\beta-1)t_{\rm hb}^{\beta}}\rangle T_{\rm obs}^{\beta-1}-\langle C_{\rm obs}\rangle\label{eq:nxvt5}).
\end{equation}

Therefore, when Eq. \ref{eq:nxvt5} is dominated by the first term, we should observe $\langle\sigma_{\rm nxv,obs}^2\rangle\propto T^{\beta-1}_{\rm obs}$ for sources with similar redshifts. If we can find a set of light curves with enough lengths, we should be able to constrain $\beta$ in this way. It should be pointed out that the deduction is similar when adopting $T_{\rm rest}$ instead of $T_{\rm obs}$, if we are only concerned about constraining $\beta$ using samples with small redshift ranges.

Based on the observations, we divide the light curves in the 7~Ms \cdfs into 4 segments, whose lengths are $\sim 4\times10^6$~s, $\sim 1\times10^7$~s, $\sim 3\times10^7$~s, and $\sim 4\times10^7$~s in the observed frame, as shown in Figure \ref{fig:lcsamp}. We perform tests similar to Section  \ref{sec:nxv} to obtain light curve samples not biased by low counts. Furthermore, these 4 segments are all unevenly sampled light curves, therefore we also perform similar simulations to quantify the bias factor $b$ as in Section~\ref{sec:psdmod}. It should be noted that we do not use other types of light curves such as the combination of 2 or 3 epochs to prevent using a segment repeatedly, so that each point in the $\sigma_{\rm nxv,corr}^2-T_{\rm obs}$ relation is based on an independent measurement.

\begin{table}[tp]
    \begin{center}
        \caption{$\sigma_{\rm nxv,corr}^2-T_{\rm obs}$ Fitting Results}\label{tab:tnxv}
        \resizebox{\linewidth}{!}{
        \begin{tabular}{ccccc}\hline\hline
            Sample& $z$ & $L_{\rm 2-10~keV}$ ($10^{42}~\rm erg~s^{-1}$) & $a$ & Const \\\hline
            A& 0.5--1& $>2.8$ & $0.53\pm0.19$ & $-5.1\pm1.5$\\
            B& 1--1.5 & $>7.6$ & $0.34\pm0.09$ & $-3.7\pm0.6$\\
            C& 1.5--2.5 & $>27$ & $0.38\pm0.15$ & $-4.2\pm1.1$\\
            D& 2.5--3.5 & $>60$ & $0.41\pm1.0$ & $-5.0\pm7.1$\\ \hline
        \end{tabular}
        }
    \end{center}
\end{table}

\begin{figure}[tp]
    \includegraphics[width=1\linewidth]{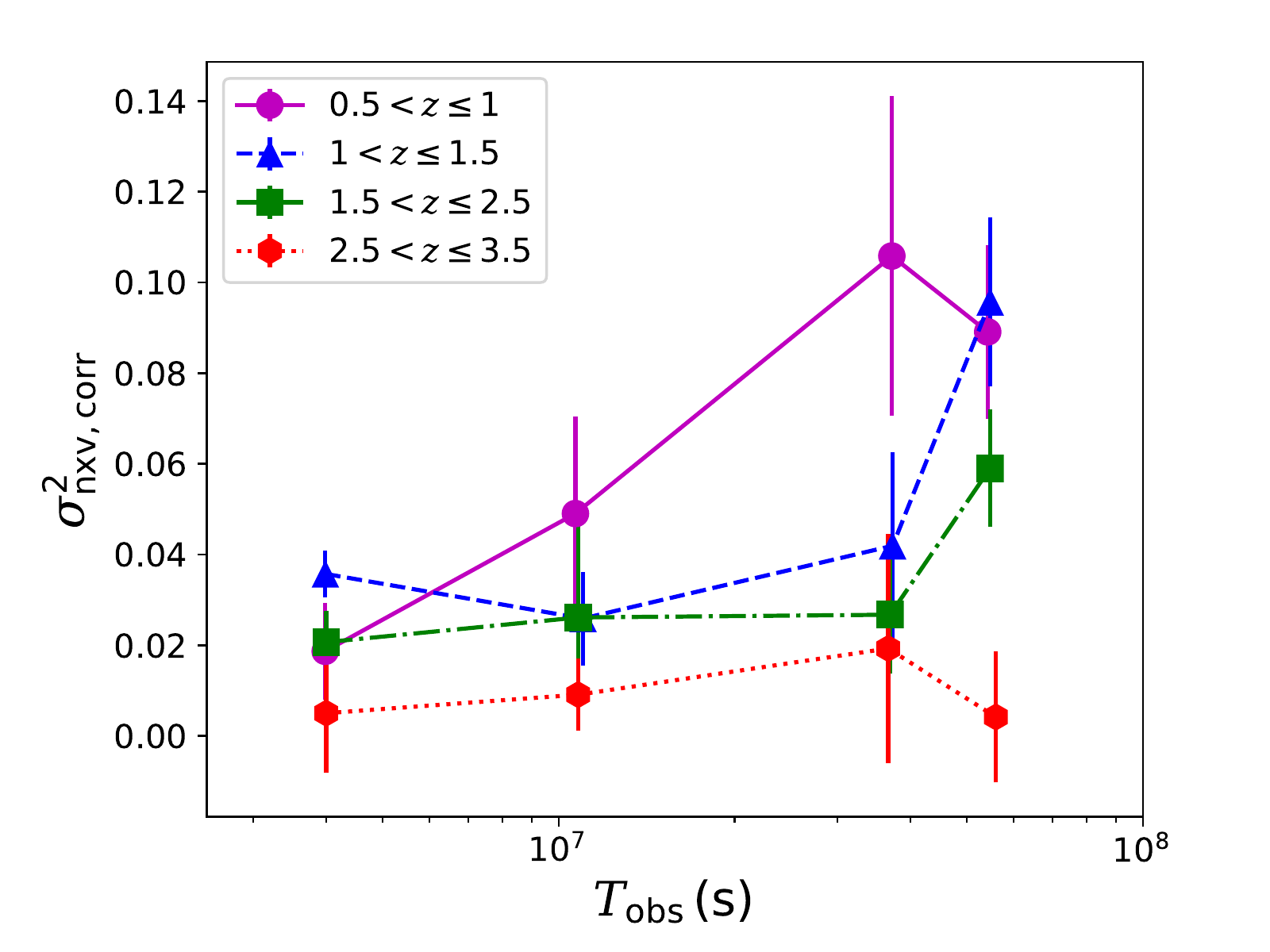}
    \caption{Plot of $\sigma_{\rm nxv,corr}^2$ vs. $T_{\rm obs}$. To avoid the bias due to sample incompleteness, we adopt the four complete subsamples indicated in Fig.~\ref{fig:csample} that have different redshift ranges and proper luminosity ranges. Each point is binned by 10--30 sources with similar time lengths in the observed frame. In the binning process, outliers beyond the 3$\sigma$ range of other data are abandoned. As expected, we could find an overall increasing trend.}\label{fig:t_nxv}
\end{figure}

\begin{figure}[tp]
    \vspace{2mm}
    \includegraphics[width=\linewidth]{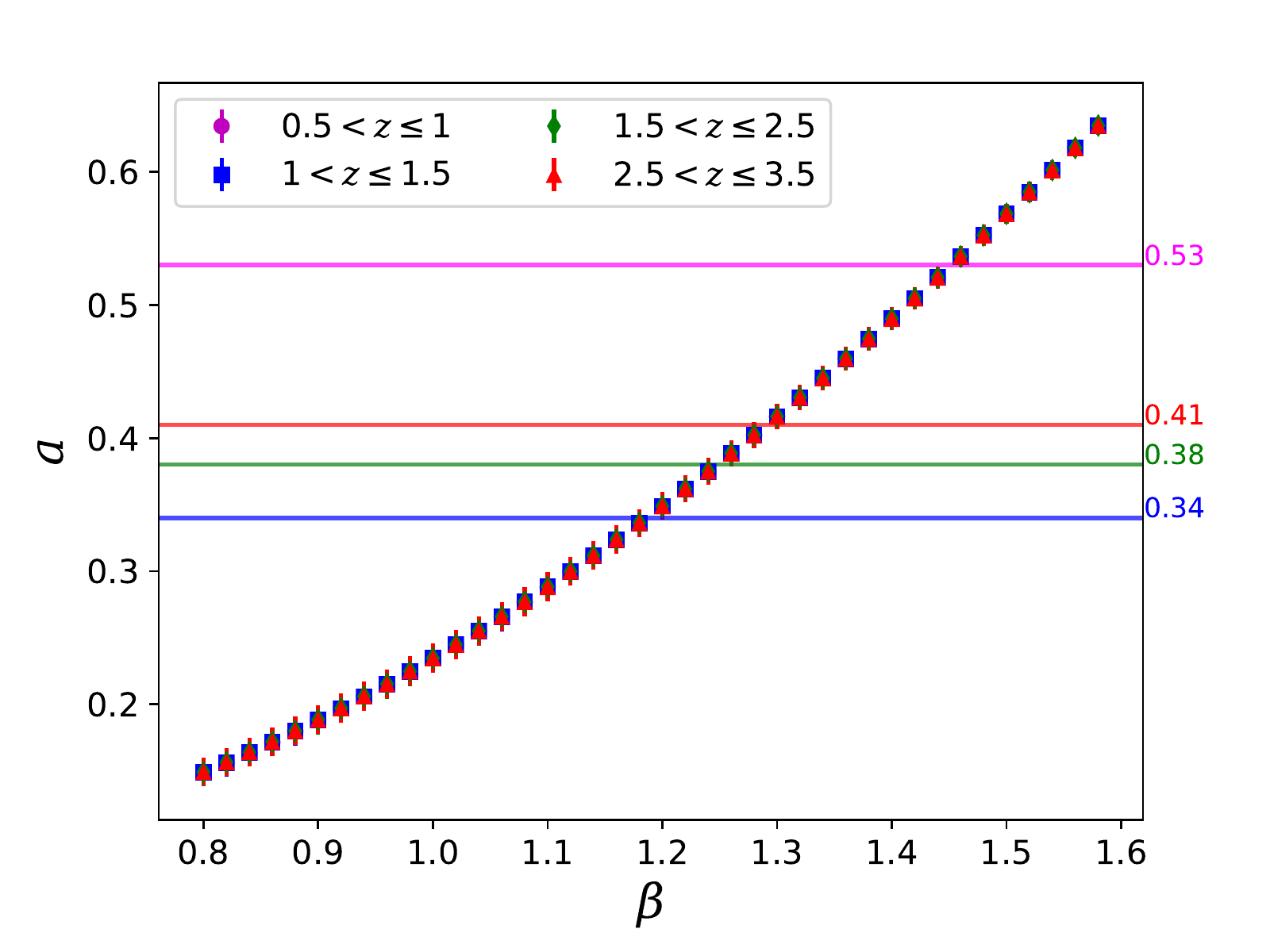}
    \caption{$a-\beta$ relation from simulation. PSD shape parameters are randomly selected in proper ranges. Four colors denote different redshift ranges in the simulation. $a-\beta$ relations are consistent with each other despite of the different redshift ranges we set. We also mark the positions of best-fit $a$ (see Table~\ref{tab:tnxv}) in the plot.}
    \label{fig:abeta}
\end{figure}

\begin{figure*}[tp]
\includegraphics[width=\linewidth]{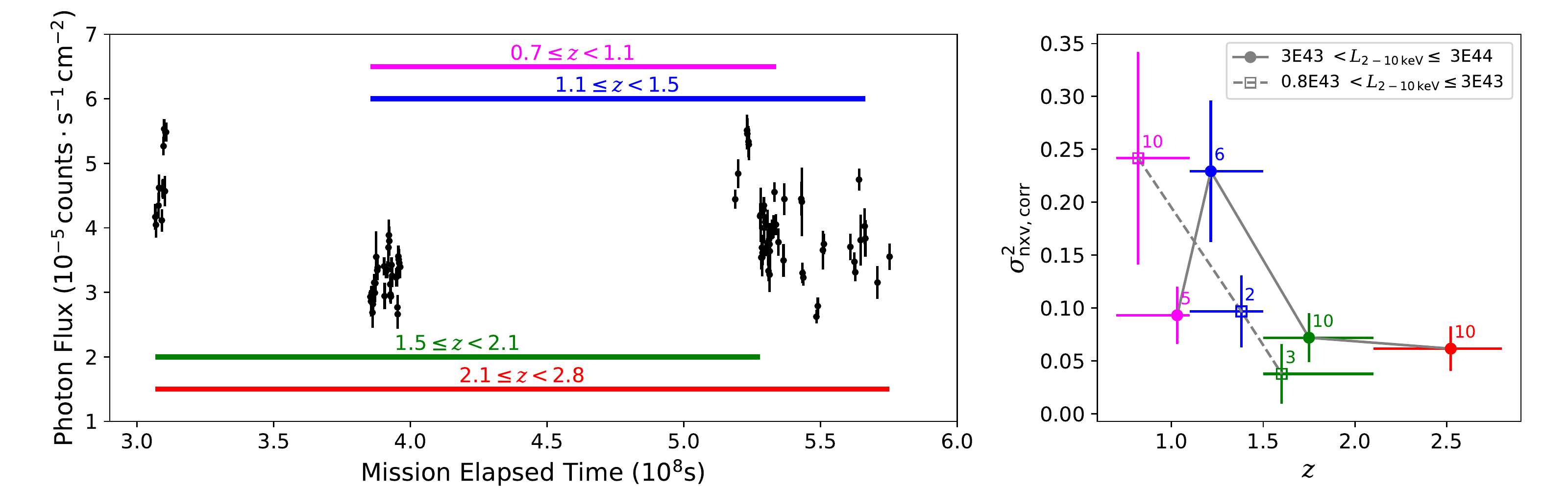}
\caption{(Left) Partial light curve of the source with XID=495, as an example to show the segments we adopt in the $\sigma_{\rm nxv,corr}^2$ computation.
Such a division scheme ensures that light curves of sources in different redshift ranges have roughly same rest-frame time lengths. 
The first 1~Ms observations (i.e., Epoch~I) are not shown. 
(Right) $\sigma_{\rm nxv,corr}^2$ vs. $z$. The position of each point is determined by the mean value of $\sigma_{\rm nxv,corr}^2$ and the median redshift. Different symbol shapes stand for different source luminosity ranges. Redshift error bars stand for the redshift ranges (see the left panel), while $\sigma_{\rm nxv,corr}^2$ error bars are standard errors (for more than three sources within a redshift bin) or based on the Vaughan et~al. (2003) estimation (i.e., Eq.\ref{eq:verr}). The amount of sources in each bin is annotated.}
\label{fig:nxv2z}
\end{figure*}

We use the four complete subsamples indicated in Fig.~\ref{fig:csample} to plot the $\sigma_{\rm nxv,corr}^2-T_{\rm obs}$ relation in Fig.~\ref{fig:t_nxv}. In the binning process, we exclude outliers with \nxv beyond the 3$\sigma$ range of other sources in each bin because we find the \nxv measurements of these outliers (1 or 2 at most in each bin) usually suffer from 1 or 2 points in the light curves with abnormally large values due to large errors or bursts. This effect is negligible for the 102-point light curves but severely influences short light curves. As a result, it is not surprising for us to find an increasing trend in most of the subsamples except for the highest-redshift one (i.e., Sample~D). Note that we only have 4 data points and the highest-redshift subsample is the smallest one (each point is binned by 10 to 15 sources).

We use a power law model to fit the $\sigma_{\rm nxv,corr}^2-T_{\rm obs}$ relation,
\begin{equation}
    \mathrm{log}~ \sigma_{\rm nxv,corr}^2 = a~ \mathrm{log} ~T_{\rm obs} + Const.\label{eq:tnxv}
\end{equation}
The fitting results are listed in Table \ref{tab:tnxv}. Theoretically, the slope $a$ in Eq.\ref{eq:tnxv} and the low-frequency slope of PSD $\beta$ are connected in the form of $a\sim \beta-1$ if $T_{\rm obs}$ is long enough, but obviously, our light curves are not ideal. For a light curve with a length of $10^7$~s, a bin size $\Delta t_{\rm obs}$ of $80$~ks, and originating from a PSD with a break frequency of $\sim 10^5$, the first term of Eq. \ref{eq:nxvt3} is only about 2 times larger than the second term. Furthermore, the difference in source properties can also introduce bias. 

To find out the exact dependence between $a$ and $\beta$, we perform a simple simulation. We use a broken power law PSD model, and randomly select 100 sets of redshifts and PSD shape parameters (normalization and high-frequency break). Through Eq.\ref{eq:bint} we obtain the expected $\sigma_{\rm nxv,corr}^2$ in 4 timescales. Then we fit the expected $\sigma_{\rm nxv,corr}^2$--$T_{\rm obs}$ relation, and find the value of $a$ we will obtain when we use different $\beta$. We present the result in Figure \ref{fig:abeta}. 

We find the corresponding $\beta$ values in Fig.~\ref{fig:abeta} for the two low-redshift subsamples (i.e., Samples A and B) are likely to be $\approx 1.2$--1.4, consistent with that derived from the $L_{\rm X}-\sigma^2_{\rm nxv}$ relation. We stress that this $\beta-a$ relation only assumes a broken power law PSD and does not depend on a specific model in Section~\ref{sec:psdmod}. The consistency between the results obtained in these two different ways proves the reliability of our $\beta$ estimation.   
For the two high-redshift subsamples (i.e., Samples C and D), the upward trends in Fig.~\ref{fig:t_nxv} are not significant, which should be due to that their intrinsic variability is weak and the rest-frame lengths of light curves are short, leading to a weak trend. Moreover, the small number of sources can also be a problem. In this case, we decide to draw our conclusion based on the low-redshift results.

\subsection{Variability evolution}\label{sec:var2z}
We have known that by PSD model fitting it is not enough to tell whether AGN variability changes in different cosmic eras. A direct way to explore this question is to compare \nxv of sources from different redshift ranges. However, from Eq. \ref{eq:bint} and Section \ref{res:Lx}, we also know that the measured \nxv suffers from the differences of luminosity ranges, rest-frame timescales, and sampling patterns of different redshift samples.

In order to reduce the influence of luminosity differences, we select a complete luminosity-limited subsample. This subsample only contains AGNs with $3\times 10^{43}~\rm erg~s^{-1}$ $<L_{\rm 2-10~keV}\leq 3\times10^{44}~\rm erg~s^{-1}$ and $0.7<z\leq2.8$ . We aim to compare the variable amplitudes (i.e., \nxv of light curves with same rest-frame lengths) of different redshift subsamples. 

Since $t_{\rm rest}=t_{\rm obs}/(1+z)$, we choose four light curve segments corresponding to four representative redshifts $z$=0.9, 1.3, 1.8, and 2.4, which are noted with horizontal lines in the left panel of Fig.~\ref{fig:nxv2z}, so that the rest-frame lengths of light curves $t_{\rm rest}$ are consistent ($t_{\rm rest}\simeq 8\times10^{7}~\rm s$), making \nxv from different redshift subsamples straightforwardly comparable. We also choose proper redshift bins (also noted in the left panel of Fig.~\ref{fig:nxv2z}) in source selection to make the variation of $t_{\rm rest}$ within each bin less than 10\%. These light-curve segments are the best choices available to make use of the longest light curves possible and ensure the consistency of the rest-frame timescales of all sources. The bias from irregular sampling pattern is also determined by simulation mentioned in Section \ref{sec:psdmod}. After all these adjustments and bias corrections, we choose the sources with photon flux $>4\times10^{-7}~\rm counts~s^{-1}~cm^{-2}$, which corresponds to the threshold of $>300$~counts for reliable \nxv measurement (see Section~\ref{sec:finsamp}), and then obtain the non-biased $\sigma_{\rm nxv,corr}^2$-$z$ relation. The result is plotted in the right panel of Fig.~\ref{fig:nxv2z}. 

Due to the above strict source-selection criteria, the available source numbers in the four redshift bins are only 5, 6, 10, and 10, respectively. According to the requirement suggested by \citet{Allevato13}, it is difficult to draw any reliable conclusion with these small bin sizes. 
Therefore, we cannot reach a definitive conclusion about whether there is an evolution of variability, and only list below some intriguing hints from the results.

Firstly, the $\sigma_{\rm nxv,corr}^2$--$z$ relation displays an overall decreasing trend. 
If it is a real trend, it could be due to the changing of PSD shape rather than accretion rate (see Fig.~\ref{fig:lxmod}),
because both our PSD fitting results based on Model~1 or 2 and other studies \citep[e.g.,][]{McLure04,Paolillo04,Paolillo17,Papadakis08} infer smaller or constant Eddington ratios toward lower redshifts, which appears contrary to what this observed evolution shows. 
Alternatively, the likely energy-band dependence mentioned in Section~\ref{res:bands} can be another potential possibility.

%

Secondly, there appears a peak at $z\simeq 1.3$ atop the overall decreasing trend. 
If it is a real feature, it is unlikely to be connected with large-scale structures (LSSs) in the E-CDF-S \citep[e.g.,][]{Gilli03,Treister09,Silverman10,Dehghan14,Xue17} since LSSs do not exist only around this redshift. 
\citet{Paolillo17} also ruled out this possibility in a relevant analysis.
Interestingly, we notice that \citet{Ueda14} found a peak of X-ray emissivity for AGNs with $\mathrm{log}~L_{\rm 2-10~keV}=43-44$ (see Fig.~20 in that work) that is close to the peak here. 
We then repeat our procedure to plot the $\sigma_{\rm nxv,corr}^2$--$z$ relation for sources with $L_{\rm 2-10~keV}=8\times 10^{42}-3\times 10^{43}~\rm erg~s^{-1}$ and $z=0.7-2.1$ in the right panel of Fig.~\ref{fig:nxv2z}.
The trend becomes monotonically decreasing, which also seems to be in line with the peak shifting behavior of AGN X-ray emissivity shown in \citet{Ueda14}.
%

\section{Transient events}\label{sec:trans}
\subsection{Event searching}

We utilize the 7~Ms \cdfs data to search for likely transient events, especially TDEs, and then constrain their occurrence rate. For this purpose, it is not appropriate to only consider the sources in L17. Since the L17 source detection is based on average fluxes over the 7~Ms timespan, it is possible that some sources lying below the nominal detection limits (thus not included in L17) may become detectable when an outburst occurs. Actually, most TDEs were found in non-active galaxies that are usually not very bright in X-rays. Therefore we also take into account the galaxy sample described in \citet{Xue10}. This sample contains 100,318 galaxies in the E-CDF-S field \citep{Xue16} and the vast majority of them have redshift and stellar-mass estimates (thus, masses of potential central black holes can be roughly estimated based on the galaxy-SMBH mass scaling). Not all of these galaxies are adopted because some of them are too faint to be detected even if the central black hole is accreting at the Eddington limit level, and the central black hole masses in some galaxies do not satisfy the requirement for a TDE \citep{Frank76, Luo08b}. As a result, the sources in the final galaxy sample considered should meet all the following criteria:

\begin{itemize}
\item[a.] The stellar mass is between $2\times10^{7}\rm M_{\odot}$ and $1.5\times10^{11} \rm M_{\odot}$. This stellar-mass range roughly corresponds to a central black hole mass range from $1\times10^{5} \rm M_{\odot}$ to $3\times 10^{8} \rm M_{\odot}$ \citep{Luo08b} adopting a scaling factor of 200--500 between stellar mass and black hole mass (Kormendy \& Ho 2013).
\item[b.] The galaxy should have an expected full-band flux of $\gsim 1.5\times10^{-7} \rm counts~ s^{-1}~cm^{-2}$ such that it would become detectable in an outburst if its central black hole is accreting at the Eddington limit. This flux limit is derived based on the typical background fluctuation level $\sigma_{\rm bkg}$ of about $10^{-8}-10^{-7}~\rm counts~ s^{-1}~cm^{-2}$ in the 7~Ms CDF-S, assuming a $\Gamma =1.8$ power law and adopting the $k_{\rm bol}-L_{\rm Edd}$ relation from \citet{Lusso10}. 
\item[c.] The galaxy is covered by all 102 \cdfs observations, which ensures that each bin has enough exposure time.
\item[d.] The galaxy is located outside of the $R_{\rm bkg,in}$ (see Table~\ref{tab:aper}) of any sources in the 7~Ms \cdfs main catalog. 
\end{itemize}

\begin{figure}[tp]
\includegraphics[width=1\linewidth]{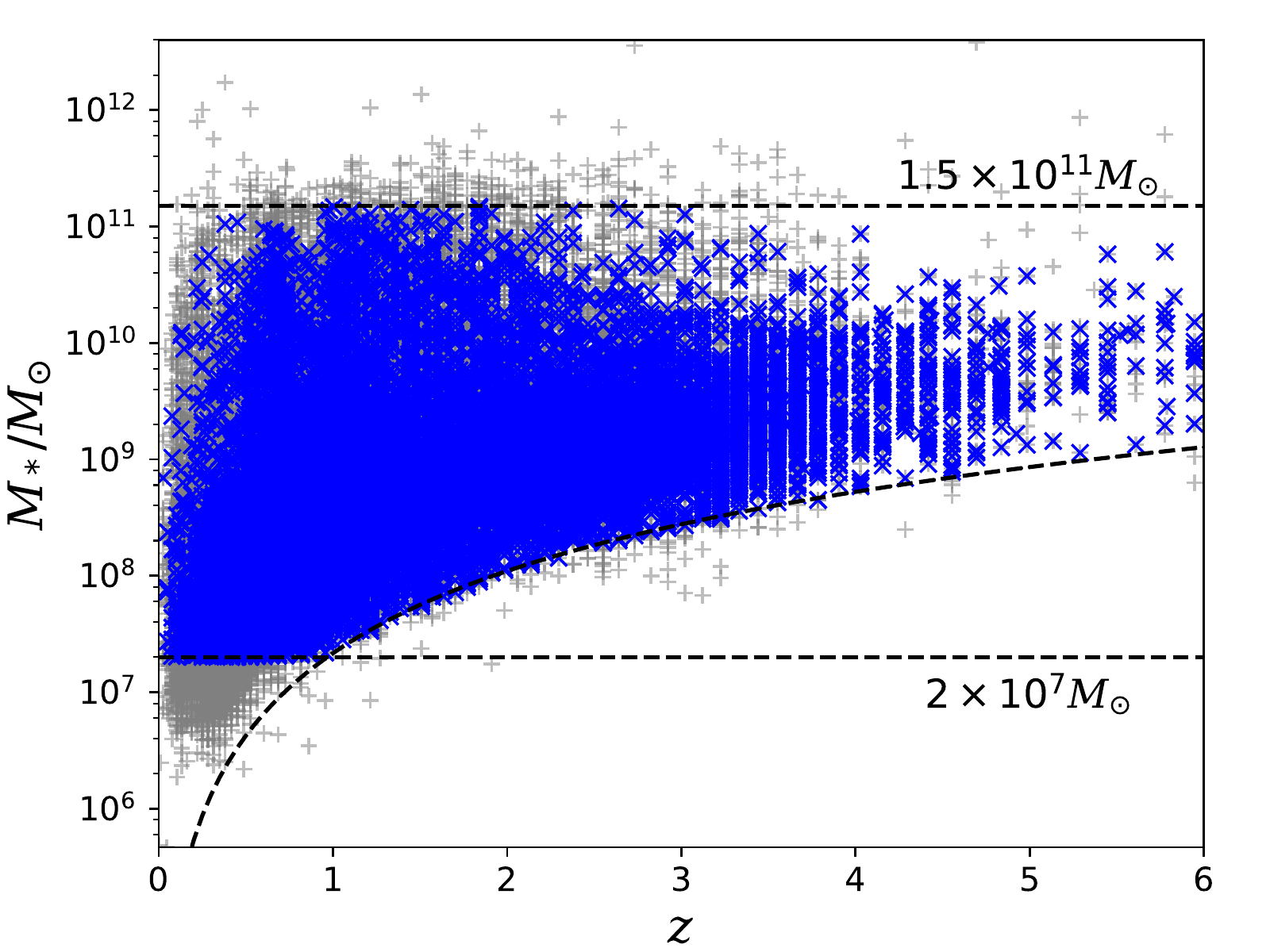}
    \caption{Redshifts and stellar masses of a large galaxy sample in the E-CDF-S. Gray crosses stand for all 100,318 galaxies in \citet{Xue10}, while blue `x' symbols denote the 19,599 non-X-ray sources in the final galaxy sample. The stellar-mass thresholds and flux limit are shown, with the latter computed assuming a $\Gamma=1.8$ power law and the \citet{Lusso10} $k_{\rm bol}-L_{\rm Edd}$ relation.}
    \label{fig:galzm}
\end{figure}

There are a total of 19,599 galaxies (without L17 detection) in the final galaxy sample, which is supplemented by the 764 L17 main-catalog X-ray sources that are covered by all 102 \cdfs observations and not in crowded X-ray source regions (e.g., pairs or triplets). The redshifts and stellar masses of the non-X-ray galaxies in the final galaxy sample are shown in Fig.~\ref{fig:galzm}.

\begin{figure*}[tp]
    \includegraphics[width=\linewidth]{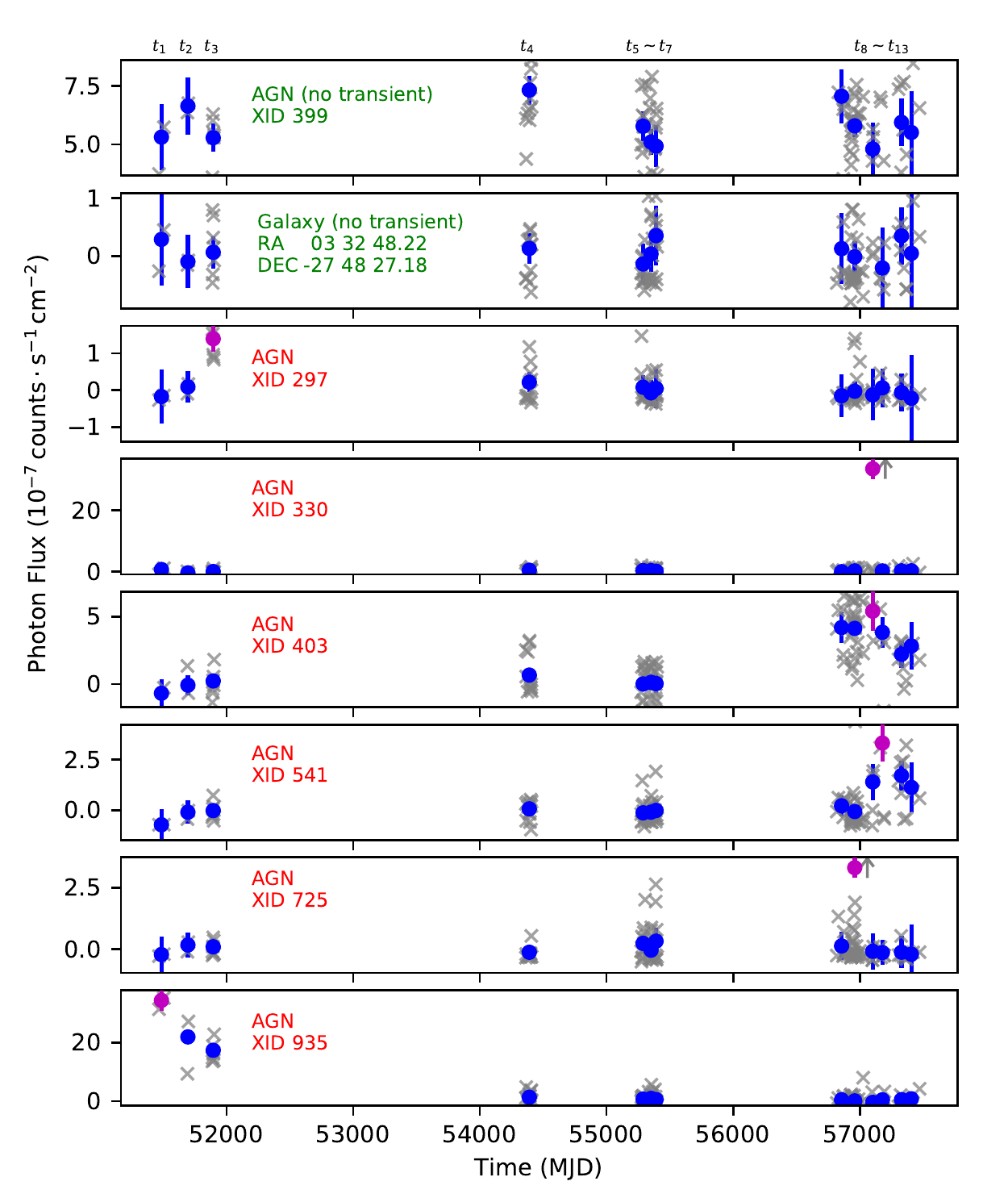}
    \caption{3-month-bin light curves of some sources. Blue dots stand for average fluxes in 3 months; magenta points denote the maximum fluxes. The time unit is the modified Julian day (MJD). Gray crosses are the fluxes binned over one individual observation. The upward arrows in the fourth and seventh panels mark the very high gray points not shown in the plot (their X-axis coordinates are slightly shifted for clarity). The first two panels are light curves of two normal sources (i.e., no transient events detected), while the other panels are light curves with candidate outbursts. The XID in the 7~Ms \cdfs main catalog or the sky position in \citet{Xue10} and the source classification in the 7~Ms main catalog are also annotated. We also mark atop $t_i$ values (t1--t13) used in Section~\ref{sec:contde}.}
    \label{fig:binlc}
\end{figure*}

Previous studies \citep[e.g.,][and references therein]{Auchettl17} have shown that TDEs usually last for a few months to a few years. Therefore, we adopt 3-month bins in our analysis to increase the S/N and avoid smoothing burst-like features. We display some binned full-band light curves with blue dots in Fig.~\ref{fig:binlc}. 

The next step is to search for transient/burst events using the 3-month-bin light curves. For each light curve, we first identify the highest flux $f_{\rm max}$ and its error $\sigma_{\rm err,m}$, and then compute the average flux $\bar{f}_{\rm normal}$ and standard error $\sigma_{\rm normal}$ of the remaining data points. If there is an outburst, the flux change $\Delta f=f_{\rm max}-\bar{f}_{\rm normal}$ should be significantly larger than the normal variability and statistical error of the source. Therefore, we select sources with $\Delta f/\sigma_{\rm normal}>3$. We also notice that short exposure times of some data points (especially the first and the last) in the light curves would cause some mis-identifications in this process. So we also require the candidates to have $\Delta f/\sigma_{\rm err,m}>3$. 

Besides the above $\Delta f$ criteria, the variable factor is taken into account as well. It is usually defined as $f_{\rm max}/f_{\rm min}$ or $f_{\rm max}/\bar{f}_{\rm normal}$ \citep[e.g.,][]{Donley02,Luo08b}. Since the existence of negative fluxes is inevitable after background subtraction, we adopt $f_{\rm max}/\bar{f}_{\rm normal}$ in our analysis. If a source satisfies at least one of the following three situations, we regard this source as a candidate transient:

\begin{itemize}
\item[ (1)] $\bar{f}_{\rm normal}>0$, $f_{\rm max}/\bar{f}_{\rm normal}>20$.
\item[ (2)] $\bar{f}_{\rm normal}\le 0$, $f_{\rm max}/ (\bar{f}_{\rm normal}+\sigma_{\rm normal})>20$.
\item[ (3)] $\bar{f}_{\rm normal}+\sigma_{\rm normal}\le 0$, $f_{\rm max}>0$.
\end{itemize}

We stress that the second and third situations are possible, because in the calculation of $\bar{f}_{\rm normal}$ the highest data point (and probably the highest few points; see the next paragraph) is not used. In this case, the signal of a galaxy may be smoothed by background fluctuation, leading to a negative $\bar{f}_{\rm normal}$ measurement.

It is possible that the transient event lasts for a very long time and we would miss it, since we will obtain elevated $\bar{f}_{\rm normal}$ and $\sigma_{\rm normal}$ by including the data points adjacent to the peak. To include such events, for sources that do not meet $\Delta f$ and the variable factor criteria, we recalculate $\bar{f}_{\rm normal}$ and $\sigma_{\rm normal}$ by excluding the highest $n (=2-6)$ data points and then check $\Delta f$ and the variable factor iteratively using the original $f_{\rm max}$ and new $\bar{f}_{\rm normal}$ and $\sigma_{\rm normal}$.
If $\Delta f$ and the variable factor are large enough after we exclude $n (\leq 6)$ data points, the source will also be considered as a candidate hosting an outburst. In contrast, a source will not be considered as a candidate with an outburst during the observations, if it cannot pass the test even after the 6 highest data points being excluded.
This step may also introduce spurious fluctuations. Therefore we perform a final visual inspection to see if the highest points are close to each other, which would be the situation for real long-duration outbursts. 

Finally, we find a total of 6 candidate transients in our galaxy sample. Basic information and light curves of these candidates are shown in Table~\ref{tab:burst} and Fig.~\ref{fig:binlc}. All these candidates are detected in L17 and satisfy our first criterion. From their light curves, these 6 sources could be roughly divided into two types. One type is long outburst. The outbursts of XID=297, XID=403, XID=541, and XID=935 last for at least a number of months and they are covered by several observations. When we inspect their 102-data point light curves, some of the candidate outbursts become less evident. Particularly, XID=403 has been reported in L17 and will be studied in depth in Wang et~al. (in prep.). The other type is short outburst, including XID=330 and XID=725. Their outbursts happened in a single observation and become extremely evident in the 102-data point light curves. When looking into these two candidates, we find that their count rates rose to $10^{-2}$ to $10^{-1}$~$\rm counts~ s^{-1}$ within just a few hundred seconds and then went back to the normal level slowly after a few thousand seconds. One of these two sources, XID=725, has also been reported in L17. More details about this source, including its likely origin, can be found in \citet{Bauer17}. We will discuss all these 6 candidates (particularly XID=330) further in a future work (Zheng et~al. in prep.).
%
Inspired by the discovery of XID=330 and XID=725, we also perform a similar test to the 102-data point light curves of both nomal galaxies and X-ray sources, but find no additional fast burst candidates.

We note that all the above 6 candidate transients are classified as AGNs in L17, which may not be appropriate. This is due to that the L17 source detection and classification are based on the entire 7~Ms \cdfs data (i.e., stacking all individual observations), which means that photons from a transient event could dominate the overall spectrum of the source, thus likely affecting the source classification.

\begin{table*}[tp]
    \begin{center}
        \caption{Sources with candidate transient events}\begin{tabular}{ccccccccc}\hline\hline
	XID$^{ (a)}$ & RA & DEC & $z^{ (b)}$ & Peak time$^{ (c)}$ & $f_{\rm max}$ ($\rm s^{-1}~cm^{-2}$)$^{ (d)}$ & $\bar{f}_{\rm normal}$ ($ \rm s^{-1}~cm^{-2}$) & $f_{\rm max}/\bar{f}_{\rm normal}$ & Type \\ \hline
	297 & 53.069719 & $-27.777204$ & $1.24^{+0.08}_{-1.17}$ & 2000/12 & $1.39\times 10^{-7}$ & $1.14\times 10^{-9}$ & 122.4 & Long \\
	330 & 53.076485 & $-27.873395$ & 0.74 & 2015/03 & $3.35\times 10^{-6}$ & $2.82\times 10^{-8}$ & 118.8 & Short \\
	403 & 53.094719 & $-27.694609$ & $1.51^{+0.03}_{-0.01}$ & 2015/03 & $5.39\times 10^{-7}$ & $2.06\times 10^{-8}$ & 26.2 & Long \\
	541 & 53.122333 & $-27.734364$ & $-1.0$ & 2015/06 & $3.33\times 10^{-7}$ & $1.09\times 10^{-8}$ & 30.4 & Long \\
	725 & 53.161561 & $-27.859342$ & $2.14^{+0.37}_{-0.56}$ & 2014/10 & $3.31\times 10^{-7}$ & $2.29\times 10^{-9}$ & 144.2 & Short \\
	935 & 53.248664 & $-27.841828$ & 0.25 & 1999/11 & $3.43\times 10^{-6}$ & $6.23\times 10^{-8}$ & 55.0 & Long \\ \hline
	\end{tabular} \label{tab:burst}
    \end{center}
    {\sc Note.} --\\
    $a$: All sources are included in the L17 7~Ms \cdfs main catalog, with their XIDs shown here.\\
    $b$: Redshifts with upper and lower errors are photometric redshifts; the value of $-1.0$ indicates no reliable redshift measurement available; and the remaining are spectroscopic redshifts.\\
    $c$: This is the time when a source reached its highest flux level. The time values are directly read from the 3-month-bin light curves, thus being not very accurate. However, accurate outburst times for the short outbursts XID=330 and XID=725 could be determined (see the text for details).\\
    $d$: The maximum fluxes are also derived from the 3-month-bin light curves, i.e., being the mean values over 3 months. For the short outbursts XID=330 and XID=725, their maximum fluxes calculated from the 102-data point light curves are much higher than the values quoted here. \\
\end{table*}

\subsection{Constraining TDE rate}\label{sec:contde} 

With the results of candidate transient event searching, we can make a rough estimation of the TDE rate $\dot{N}_{\rm TDE}$ in our sample. Based on the algorithm outlined in \citet{Luo08b},  we first compute the total rest-frame time $T_{\rm total}$ we inspect:
\begin{equation}
T_{\rm total} = \sum_{i}^{N_{\rm src}} T_{i,\rm eff}/ (1+z_i). 
\end{equation} 
$T_{i,\rm eff}$ is the effective exposure time of the $i$th source in the observed frame. $N_{\rm src}$ and $z_i$ are the amount of our sources and their redshifts. For sources without any redshift estimates (only 14 sources), we assign them the median redshift of our sample $z_{\rm med}=1.27$. Two situations should be considered separately: long and short outbursts, with the former likely being TDEs. 

For long outbursts, considering that they could last for several months, we should still be able to detect such outbursts if they occur a few months ahead of each of the four epochs (see Fig.~\ref{fig:binlc}). If we only consider outbursts that can be detected in 3 months (rest-frame time), $T_{i,\rm eff}$ can be estimated by
 \begin{eqnarray}
 T_{i,\rm eff} &=& 3 ~\mathrm{months}\times (1+z_i)\times 4 \nonumber \\
 		& &+ (t_3-t_1)+(t_7-t_5)+(t_{13}-t_8).
 \end{eqnarray}
 Here $t_i$ is the time of the $i$th data point in the 3-month-bin light curve (see Fig.~\ref{fig:binlc}). 
 
Assuming that any outburst occurring during the observations would be detected, we could obtain the event rate using
\begin{equation}
\langle\dot{N}_{\rm event}\rangle = \frac{N_{\rm event}}{T_{\rm total}}\rm galaxy^{-1}~ yr^{-1}.
\end{equation}
According to Gehrels (1986), we can derive the 90\% confidence-level upper limit and lower limit of the amount of transient events $N_{\rm event}$. For long outbursts, if all candidates were associated with TDEs, we would have $\langle\dot{N}_{\rm TDE}\rangle=8.6^{+8.5}_{-4.9}\times 10^{-5}~\rm galaxy^{-1}~ yr^{-1}$. 
Our $\langle\dot{N}_{\rm TDE}\rangle$ estimation is consistent with other studies. Previous observational results \citep[e.g.,][]{Donley02,Luo08b,van14} found TDE rates in their studied samples to be $10^{-6}-10^{-4}~\rm galaxy^{-1}~yr^{-1}$, while theoretical studies \citep[e.g.,][]{Wang04,Stone16} indicated $\dot{N}_{\rm TDE}=10^{-5}-10^{-3}~\rm galaxy^{-1}~ yr^{-1}$. As mentioned before, our $\langle\dot{N}_{\rm TDE}\rangle$ calculation is crude. Uncertainties may be introduced due to sample selection, detection efficiency, and some other issues. We briefly introduce the influences of these issues below.

The first issue is from sample selection. 
Our sample is flux limited and has a stellar-mass range of $2\times10^{7}~\rm M_{\odot}$--$1.5\times10^{11}~\rm M_{\odot}$ (see Fig.~\ref{fig:galzm}).
First, previous studies (e.g., Wang \& Merrit 2004; Stone \& Metzger 2016) pointed out that $\dot{N}_{\rm TDE}$ is anti-correlated with black hole mass, which indicates that our estimated average TDE rate is likely to be slightly underestimated, given that massive galaxies make up a larger fraction in our sample than in a complete sample.
Second, our sample volume might be overestimated (thus the TDE rate being underestimated) with the adopted broad stellar-mass range,
which originates from the galaxy-SMBH mass scaling relation that has large scatters and uncertainties (e.g., we use a scaling factor of 500 to estimate the upper limit of stellar mass and 200 to estimate the lower limit).  
Third, a relevant point is that, if a scaling factor around 1000 (suggested by, e.g., \citealt{Haring04} and \citealt{Sun15}) was adopted, additional very massive galaxies would be included in our sample, but this increase of our sample volume would be less than 0.1\% given the scarcity of such galaxies.
Fourth, we choose a scaling factor of 200 to estimate the Eddington luminosity of central black hole, which would overestimate the black hole mass and maximum flux and thus include some sources that are not able to be detected even in the outburst state, resulting in the overestimation of our sample volume.
Finally, we choose a uniform flux limit in source selection, as opposed to the fact that
the X-ray flux limit varies significantly across the \cdfs field of view \citep[e.g.,][]{Xue11,Luo17}.
Therefore, it is possible that we would have been able to detect some outbursts from galaxies not included in our galaxy sample (due to the flux-limit cut), especially for galaxies near the central field of view where the flux limit is much smaller than the adopted value.
This means an underestimation of our sample volume, thus leading to an overestimate of the TDE rate.
However, given that there is an anti-correlation between $\dot{N}_{\rm TDE}$ and black hole mass (see the first point above) and that these ``missed'' galaxies tend to be less massive than the sources in our sample,
the inclusion of these ``missed'' galaxies into our sample should boost the estimated TDE rate. 
In fact, the peak flux of the outburst candidate XID=297, with a small off-axis angle of $\approx 3\arcmin$, is below the flux limit we set for source selection (see Table~\ref{tab:burst}). 
The above various factors bring some uncertainties to the estimate of the TDE rate, most of which, if treated properly, tend to increase the estimated TDE rate.

Detection efficiency is another important issue that would influence the estimate of TDE rate. In the searching, we assume that all outbursts occurring during the exposures could be found, no matter what their properties (e.g., $N_{\rm H}$, $\Gamma$, and off-axis angle) might be. But obviously this is too ideal. A more realistic calculation should be
\begin{equation}
N_{\rm TDE} = \sum_{i}^{n} \epsilon_i \frac{T_{i,\rm eff}}{ (1+z_i)}\langle\dot{N}_{\rm TDE}\rangle
\end{equation}
\\
In this expression, detection efficiency $\epsilon_i$ should be less than 1. Therefore, the real average TDE rate $\langle\dot{N}_{\rm TDE}\rangle$ is again underestimated.

The light curve profile of an outburst also plays an important role. We estimate $\langle\dot{N}_{\rm TDE}\rangle$ assuming that outbursts are only detectable in 3 months. But from the results, we also find 3 out of the 4 long outbursts (except XID=297) are likely to be recognized in more than 1 data points in the 3-month-bin light curves. The variety of outburst profiles brings difficulty to estimate $T_{i,\rm eff}$. For longer outbursts, real $T_{\rm total}$ should be longer, leading to a smaller $\langle\dot{N}_{\rm TDE}\rangle$ than we obtain. For example, with our calculation, if we assume all outbursts could be detected in 1 year, $\langle\dot{N}_{\rm TDE}\rangle$ will become $3.4^{+3.4}_{-1.9}\times 10^{-5}~\rm galaxy^{-1}~ yr^{-1}$. To solve this problem, it is necessary to know about the intrinsic distribution of X-ray outburst durations, which is not feasible by now. 

Last but not least, further studies (e.g., Zheng et al. in prep.) are needed to confirm the nature of these outbursts. Since there are only 4 long outbursts, any mistake in classification will change the result significantly. 

For short outbursts such as XID=330 and XID=725, $T_{i,\rm eff}$ should be the true exposure time. Because we only use sources fully covered by all 102 observations, $T_{i,\rm eff}$ becomes a constant of $\approx 7.0\times 10^6~\rm s\approx 0.225~yr$. Similarly, we can obtain the frequency of this type of events $\langle\dot{N}\rangle=1.0^{+1.1}_{-0.7}\times 10^{-3}~\rm galaxy^{-1}~yr^{-1}$. Since we still lack understanding about their nature \citep[see][for detailed discussions of XID=725]{Bauer17}, we cannot assess how sample selection would influence the result. But because of their short durations and high variable factors, we think detection efficiency and light curve profile should not have important effects on estimating the event rate of short outbursts. 

\section{Summary}\label{sec:sum}

We use the 7~Ms CDF-S, the deepest X-ray survey to date, to study AGN variability across an X-ray luminosity range of $10^{41}-10^{45}~\rm erg~s^{-1}$ and a redshift range of $0-5$. Benefiting from the long monitoring timespan and exposures as well as considered analyses, we are able to obtain a number of notable results as listed below.
\begin{itemize}
\item[1.] We perform simulations to inspect the uncertainty and bias introduced by low photon counts to estimation of normalized excess variance (\nxvt). \nxv measurements would have unacceptably large scatters when sources have less than 300 counts. Therefore, we choose 300 counts as the threshold and select 148 AGNs with reliable full-band \nxv and 77 with reliable soft- and hard-band \nxv values (see Section \ref{sec:anal}).
\item[2.] We find that long-term variability is largely consistent between different energy bands for subsamples with different redshifts. This result suggests that the effect of likely variability dependence on energy band is not significant in the subsequent analysis of the $L_{\rm 2-10~ keV}-\sigma_{\rm nxv}^2$ relation for subsamples with different redshifts.
\item[3.] The similarity between the \nxv distributions of obscured ($N_{\rm H}>10^{23}~\rm cm^{-2}$) and less obscured ($N_{\rm H}\leq 10^{23}~\rm cm^{-2}$) AGNs is suggested by a K-S test. Except for a slight discrepancy most probably caused by the $L_{\rm 2-10~ keV}-$\nxv relation, the two subsamples show good consistency. The Spearman's ranking tests with the 4 complete subsamples further demonstrate that column density may not be an important factor for variability (see Section \ref{res:NH}).
\item[4.] Confirming previous studies, we find a strong anti-correlation between $L_{\rm 2-10~ keV}$ and \nxvt. We show that this anti-correlation is sensitive to Eddington ratio $\lambda_{\mathrm{Edd}}$ and the low-frequency power law index $\beta$ of AGN PSD. Using a MCMC method, we fit $L_{\rm 2-10~keV}-$\nxv relation with 4 different PSD models. Best-fit results indicate a $\beta$ of $1.2-1.3$ for all models. Results disfavor Models 3 and 4 since they require an Eddington ratio of $\lambda_{\rm Edd}\geq 1$. Fitting results (i.e., $\lambda_{\mathrm{Edd}}$ and $\beta$) of subsamples of different redshifts do not show significant differences compared to their error bars (see Section \ref{res:Lx} and Section \ref{sec:psdmod}). 
\item[5.] We investigate the $\sigma_{\rm nxv}^2-T_{\rm observed}$ relation and find an overall increasing trend. Despite of uncertainties, we also obtain $\beta \sim 1.3$ by fitting the $\sigma_{\rm nxv}^2-T_{\rm observed}$ relation, which is a model-independent method (see Section \ref{sec:tnxv}).
\item[6.] After controlling the luminosity range and the rest-frame length of light curves, we build a small yet complete sample to examine the redshift evolution of AGN variability. We reach no definitive conclusion due to limited source statistics in each redshift bin, albeit with a likely hint of decreasing AGN variability at fixed luminosity toward large redshifts
(see Section~\ref{sec:var2z}).
\item[7.] We carry out a systematic search for transient events in 19,599 normal galaxies and 764 X-ray sources in the 7~Ms \cdfs using 3-month-bin light curves. Six candidate outbursts are found. Four of them have a relatively long duration of several months, while the other two have very high variable factors and last for a short duration, which are probably a new type of fast outbursts. The detailed nature of these events are left to a future study. If these four long outbursts are all associated with TDEs, we simply estimate the average TDE rate to be $\langle\dot{N}_{\rm TDE}\rangle=8.6^{+8.5}_{-4.9}\times 10^{-5}~\rm galaxy^{-1}~ yr^{-1}$. This result is comparable to previous studies. We also do this calculation for the fast outbursts and obtain an event rate of $\langle\dot{N}\rangle=1.0^{+1.1}_{-0.7}\times 10^{-3}~\rm galaxy^{-1}~yr^{-1}$ (see Section \ref{sec:trans}).
\end{itemize}

Our work provides new clues of the low-frequency part of AGN PSD (i.e., the low-frequency slope $\beta$), where there is still no much knowledge because of the lack of longterm observations. Our result of $\beta\sim 1.2$--1.3 indicates that the power of AGN longterm variability is larger than the frequently-used assumption (i.e., $\beta=1$). An index of $\beta=1$, which is found in BHBs and some AGNs, may not be able to explain the variability behavior of low-redshift low-luminosity AGNs. Our constraint on $\beta$ is consistent with the recent result from modeling AGN UV/optical variability in the SDSS Stripe 82 \citep{Guo17}. This result could help future work build a more accurate AGN PSD model and put constraints to the physical origin of AGN X-ray variability. 

In this work, we take fully into account many factors that could affect the variability measurement, including low counts, energy-band and $T_{\rm rest}$ differences caused by different redshifts, obscuration, luminosity, as well as sample incompleteness and irregular sampling, some of which are often ignored in some previous studies. Therefore, we think our analyses are robust and not influenced appreciably by most (if not all) of these biases. 

Even with the 7~Ms \cdfs data, our analyses are sometimes confronted with small numbers of sources in a limited number of bins, largely due to the insufficient sample volume. In this situation, we are still not able to draw solid conclusions about, e.g., the dependence of variability on obscuration or the likely evolution of Eddington ratio. Such a situation will be greatly improved if additional longterm deep X-ray surveys become available.

\acknowledgments
We thank the referee for careful reading and helpful comments that help improve the paper.
X.C.Z., Y.Q.X., J.Y.L., and M.Y.S. acknowledge the support from the 973 Program (2015CB857004), the National Natural Science Foundation of China (NSFC-11473026, 11421303), the CAS Frontier Science Key Research Program (QYZDJ-SSW-SLH006), and the Fundamental Research Funds for the Central Universities.
B.L. acknowledge support from the National Natural Science Foundation of
China grant 11673010 and the Ministry of Science and Technology of China
grant 2016YFA0400702.
T.M.H. acknowledges the CONICYT/ALMA funding Program in Astronomy/PCI
Project N$^\circ$:31140020. T.M.H. also acknowledges the support from
the Chinese Academy of Sciences (CAS) and the National Commission for
Scientific and Technological Research of Chile (CONICYT) through a
CAS-CONICYT Joint Postdoctoral Fellowship administered by the CAS
South America Center for Astronomy (CASSACA) in Santiago, Chile.
F.E.B. acknowledges support from CONICYT-Chile
(Basal-CATA PFB-06/2007, FONDECYT Regular 1141218),
the Ministry of Economy, Development, and Tourism's Millennium Science
Initiative through grant IC120009, awarded to The Millennium Institute
of Astrophysics, MAS.




\end{document}